\documentclass[published]{JHEP3}
\JHEP{07(2004)038}

\skip\footins = 1\bigskipamount plus 2pt minus 4pt

\newcommand{\Gammait}{{\mit\Gamma}}
\newcommand{\tr}{\mathop{\rm tr}\nolimits}
\newcommand{\Tr}{\mathop{\rm Tr}\nolimits}
\newcommand{\Pf}{\mathop{\rm Pf}\nolimits}
\newcommand{\SU}{\mathop{\rm SU}\nolimits}
\newcommand{\SO}{\mathop{\rm SO}\nolimits}
\newcommand{\U}{\mathop{\rm {}U}\nolimits}
\newcommand{\rmd}{{\rm d}}
\newcommand{\rmD}{{\rm D}}

\title{Majorana and Majorana-Weyl fermions in lattice gauge theory}

\author{Teruaki Inagaki\\
  Graduate School of Science and Engineering, Ibaraki University\\ 
  Mito 310-8512, Japan\\
  E-mail: \email{teruaki@serra.sci.ibaraki.ac.jp}}

\author{Hiroshi Suzuki\\
  Institute of Applied Beam Science, Ibaraki University\\ 
  Mito 310-8512, Japan\\
  E-mail: \email{hsuzuki@mx.ibaraki.ac.jp}}

\abstract{In various dimensional euclidean lattice gauge theories, we examine a
compatibility of the Majorana decomposition and the charge conjugation property
of lattice Dirac operators. In $8n$ and $1+8n$ dimensions, we find a difficulty
to decompose a classical lattice action of the Dirac fermion into a system of
the Majorana fermion and thus to obtain a factrized form of the Dirac
determinant. Similarly, in $2+8n$~dimensions, there is a difficulty to
decompose a classical lattice action of the Weyl fermion into a system of the
Majorana-Weyl fermion and thus to obtain a factrized form of the Weyl
determinant. Prescriptions based on the overlap formalism do not remove these
difficulties. We argue that these difficulties are reflections of the global
gauge anomaly associated to the real Weyl fermion in $8n$~dimensions. For this
reason (besides other well-known reasons), a lattice formulation of the $N=1$
super Yang-Mills theory in these dimensions is expected to be extremely
difficult to find.}

\received{June 17, 2004}
\accepted{July 16, 2004}

\begin{document} 

\section{Introduction}\label{section1}

In $d$~dimensional Minkowski spacetime, when $d=0$, 1, 2, 3, $4\bmod
8$, it is possible to define Majorana fermions by imposing the
Majorana reality condition on the Dirac fermion. Also, when
$d=2\bmod8$, it is possible to impose the Majorana condition on the
Weyl fermion to define Majorana-Weyl fermions. In these dimensions,
Majorana and Majorana-Weyl fermions form an irreducible spinor
representation of the Lorentz group and, for this reason, these
fermions are fundamental in field theories in lower and higher
dimensions, especially in supersymmetric ones. In this paper, we study
a possibility to formulate these degrees of freedom in euclidean
lattice gauge theory. We take the Wilson-Dirac
operator~\cite{Wilson:1975id} and the overlap-Dirac
operator~\cite{Neuberger:1998fp} and examine a compatibility of the
euclidean Majorana decomposition and the charge conjugation property
of these lattice Dirac operators.\footnote{We take the Wilson-Dirac
  operator because it is the simplest Dirac operator which represents
  a single Dirac fermion and the overlap-Dirac operator because of its
  virtue concerning the chiral symmetry.  A similar analysis could be
  performed for the Dirac operator in the perfect action
  approach~\cite{Hasenfratz:1998ft}.}

\pagebreak[3]

Let us first explain what we mean by Majorana and Majorana-Weyl
fermions in an euclidean theory by illustrating how the euclidean
Majorana decomposition works in unregularized continuum theory. If one
switches the lorentzian signature to the euclidean one, the Majorana
reality condition cannot be imposed in an $\SO(d)$ invariant manner,
when $d=3$, $4\bmod 8$, and the Majorana-Weyl condition cannot be
imposed when $d=2\bmod 8$. Instead imposing these conditions in
euclidean theory, one can define Majorana and Majorana-Weyl degrees of
freedom in the following way. This prescription corresponds to, in a
level of the functional integral, the prescription
of~ref.~\cite{Nicolai:1978vc}.\footnote{For a general treatment of
  spinors in euclidean theory, see
  ref.~\cite{vanNieuwenhuizen:1996tv}.} Take the Dirac fermion defined
by the action
\begin{equation}
   S=\int\rmd^dx\,\overline\psi(x)\gamma_\mu D_\mu\psi(x)\,,
\label{onexone}
\end{equation}
where $D_\mu=\partial_\mu+A_\mu^aR(T^a)$ is the covariant derivative
with respect to a gauge-group representation~$R$. For Majorana
fermions to be defined, the representation matrix $R(T^a)$ must be
real, $R(T^a)^*=R(T^a)=-R(T^a)^T$, and this will be assumed throughout
this paper when considering Majorana and Majorana-Weyl fermions. See
appendix~\ref{section7} for our convention on gamma matrices. We then
introduce the Majorana decomposition by
\begin{equation}
   \psi={1\over\sqrt2}(\chi+i\eta),\qquad
   \overline\psi={1\over\sqrt2}(\chi^TB-i\eta^TB)\,,
\label{onextwo}
\end{equation}
where $B$ denotes the charge conjugation matrix, either $B_1$ or $B_2$
in~eq.~(\ref{axten}) depending on the dimension~$d$; for $d=0$, 1, 2,
3, $4\bmod 8$, we take, $B_1$, $B_1$, $B_1$ or $B_2$, $B_1$ and~$B_2$,
respectively. From basic properties of the charge conjugation
matrices, eqs.~(\ref{axfour}), (\ref{axeleven}) and~(\ref{axsixteen}),
one finds\footnote{In this expression, $D_\mu(x,y)=D_\mu\delta^d(x-y)$
  is the kernel of the covariant derivative in position space. The
  transpose operation~$T$ acts also on position-space indices as
  $x\leftrightarrow y$.}
\begin{equation}
   [B\gamma_\mu D_\mu(x,y)]^T=-B\gamma_\mu D_\mu(x,y)\,.
\label{onexthree}
\end{equation}
Therefore, after substituting eq.~(\ref{onextwo}) into
eq.~(\ref{onexone}), one has
\begin{equation}
   S=\int\rmd^dx\,\left[{1\over2}\chi^T(x)B\gamma_\mu D_\mu\chi(x)
     +{1\over2}\eta^T(x)B\gamma_\mu D_\mu\eta(x)\right].
\label{onexfour}
\end{equation}
A system of the Dirac fermion is thus decomposed into two mutually
independent systems, each of which may be interpreted as representing
a single Majorana fermion. We refer to this type of decomposition as
the Majorana decomposition.  The functional integral over~$\chi$, for
example, then gives rise to the pfaffian~$\Pf(B\gamma_\mu D_\mu)$ that
may be regarded as the square-root of the Dirac
determinant~$\det(\gamma_\mu D_\mu)$. It is also confirmed that for
$d=5$, 6, $7\bmod8$, for which there is no Majorana fermions in the
\emph{Minkowski} spacetime, neither $B_1$ nor~$B_2$ satisfies
eq.~(\ref{onexthree}).\footnote{For $d$~dimensional Minkowski
  spacetime, when $d=2$, 3 and~$4\bmod8$, the (single) Majorana
  fermion can acquire a mass. The euclidean Majorana decomposition of
  the Dirac mass term~$m\overline\psi(x)\psi(x)$ works precisely at
  these dimensions.}

When $d$~is even, one can define the (left-handed) Weyl fermion by
setting the chirality constraint
\begin{equation}
   P_-\psi=\psi\,,\qquad\overline\psi P_+=\overline\psi\,,\qquad
   P_\pm={1\pm\gamma\over2}\,,
\label{onexfive}
\end{equation}
in eq.~(\ref{onexone}). For $d=2\bmod 8$, one may further apply the
Majorana decomposition~(\ref{onextwo}) with either use of~$B_1$
or~$B_2$, because $P_-\chi=\chi$ (or $P_-\eta=\eta$) implies
$\chi^TBP_+=\chi^TB$ (or $\eta^TBP_+=\eta^TB$) from the property of
the charge conjugation matrices~(\ref{axsix}) in these
dimensions. Again, this consistent chirality for $\chi$ (or $\eta$) is
possible only when~$d=2\bmod8$. Then $\chi$, for example, may be
regarded as representing the Majorana-Weyl fermion. The functional
integral over~$\chi$ results in the pfaffian~$\Pf(B\gamma_\mu D_\mu
P_-)$ that may be regarded as the fourth-root of the Dirac
determinant.

As we have observed, the Majorana decomposition works perfectly in
unregularized gauge theories. However, it is not obvious if it works
also with non-perturbative regularization, such as the lattice gauge
theory. Results of our analysis in this paper can be summarized as
table~\ref{tab1}. We will find, somewhat curiously, $8n$, $1+8n$
and~$2+8n$ dimensions refuse the Majorana and the Majorana-Weyl
decompositions, even if one allows a wide class of ``lattice
modifications'' of charge conjugation matrices and the chirality
projection.\footnote{The $d=1$~case is special: when the gauge-group
  representation is real, a lattice Dirac determinant is a constant
  being independent of a gauge-field configuration. The partition
  function of the Majorana fermion can be defined as the square-root
  of this constant.}
\TABULAR[t]{|c|c|c|c|c|c|c|c|c|c|c|c|}{ \hline
  $d$&2&3&4&5&6&7&8&9&10&11&12\\ \hline\hline Dirac
  &$\bigcirc$&$\bigcirc$&$\bigcirc$&$\bigcirc$&$\bigcirc$&$\bigcirc$
  &$\bigcirc$&$\bigcirc$&$\bigcirc$&$\bigcirc$&$\bigcirc$\\ \hline
  Majorana &$\bigcirc$&$\bigcirc$&$\bigcirc$&---&---&---
  &$\times$&$\times$&$\bigcirc$&$\bigcirc$&$\bigcirc$\\ \hline Weyl
  &$\bigtriangleup$&---&$\bigtriangleup$&---&$\bigtriangleup$&---
  &$\bigtriangleup$&---&$\bigtriangleup$&---&$\bigtriangleup$\\ \hline
  Majorana-Weyl
  &$\times$&---&---&---&---&---&---&---&$\times$&---&---\\ \hline}
	{Euclidean lattice formulation of a single fermion in various
	  dimensions~$d$.  We are not interested in those entries
	  filled by~---. Entries with~$\bigcirc$ can be formulated
	  quantum mechanically as well as classically. Entries
	  with~$\bigtriangleup$ can be formulated at least
	  classically; a consistency in the quantum level must be
	  examined separately. For entries with~$\times$, we could not
	  find even a ``classical level'' formulation. The pattern
	  repeats for $d$ modulo~8.\label{tab1}}  
Phenomenologically, what we will find is a conflict between the
Majorana decomposition and the charge conjugation property of lattice
Dirac operators.  An analogous phenomenon has been known in
4~dimensional euclidean lattice
theories~\cite{Fujikawa:2001ka,Fujikawa:2002is} as a conflict between
chiral invariant Yukawa couplings and the Majorana decomposition. The
fact that CP symmetry is not
manifest~\cite{Hasenfratz:2001bz,Fujikawa:2002vj} in 4~dimensional
lattice chiral gauge theories is also related to this conflict.

The overlap-Dirac operator was originally
discovered~\cite{Kikukawa:1997qh,Neuberger:1998fp} from the Dirac
determinant obtained by the overlap
formalism~\cite{Narayanan:1993wx}. In the context of the latter
formalism, a definition of Majorana and Majorana-Weyl fermions has
been studied in ref.~\cite{Huet:1996pw} and in
ref.~\cite{Maru:1997kh}. See also
refs.~\cite{Kitsunezaki:1997iu,Kikukawa:1997qh}. According to these
analyses, many-body hamiltonians, from which lowest-energy states in
the overlap formalism are defined, are decomposed into two mutually
independent systems; this is analogous to the Majorana
decomposition. We will illustrate, however, this decomposition itself
does not imply a factorized form (such as the pfaffian) of the Dirac and the Weyl
determinants obtained by the overlap
formalism. The overlap formalism does not remove the above
difficulties.

\looseness=-1 Are these difficulties in $8n$, $1+8n$ and~$2+8n$ dimensions merely
technical ones peculiar to lattice Dirac operators that we have
studied? We do not think so. We will argue that these difficulties are
reflections of the global gauge
anomaly~\cite{Witten:fp,Elitzur:1984kr} associated to Weyl fermions
belonging to a \emph{real} representation in $8n$~dimensional gauge
theories. This is interesting because an elementary kinematical
analysis in lattice formulation (that we will do) indicates a subtle
phenomenon in quantum theory, just like the species doubling can be
regarded as a reflection of the chiral anomaly.  Namely, in the
lattice formulation, classical theory and quantum theory are
indistinguishable to some extent. On the other hand, this argument
suggests that it is extremely difficult to avoid these difficulties of
the Majorana and the Majorana-Weyl decompositions in these dimensions,
because they have a physical origin.

This paper is organized as follows. In section~\ref{section2}, the
Majorana decomposition in even dimensional spaces is considered. We
find that $8n$~dimensions refuse the Majorana decomposition, even if a
wide class of ``lattice modifications'' of the charge conjugation
matrix is allowed. In section~\ref{section3}, the Majorana
decomposition of Weyl fermions is analyzed. In these sections, the
overlap-Dirac operator is taken because, being a solution to the
Ginsparg--Wilson relation~\cite{Ginsparg:1982bj}, it has a nice
property concerning the chiral symmetry. Weyl fermions on the lattice
cannot be defined even classically without this lattice chiral
symmetry~\cite{Hasenfratz:1998ft,Luscher:1998pq,Narayanan:1998uu}. We
find that a consistent Majorana decomposition of Weyl fermions is
impossible even a wide class of lattice modifications of the chirality
projection and the charge conjugation matrix is allowed. In
section~\ref{section4}, we examine somewhat in detail a prescription
for Majorana-Weyl fermions in the overlap formalism
of~ref.~\cite{Huet:1996pw}. In section~\ref{section5}, the Majorana
decomposition in odd dimensional spaces is considered. We do not find
a successful Majorana decomposition in $1+8n$~dimensions. In
section~\ref{section6}, we present an argument that difficulties of
the Majorana and the Majorana-Weyl decompositions in $8n$, $1+8n$
and~$2+8n$ dimensions have an intrinsic meaning as a reflection of the
global gauge anomaly. Appendix~\ref{section7} is for our conventions
for gamma matrices.  Charge conjugation properties of the Wilson-Dirac
and the overlap-Dirac operators are summarized in
appendix~\ref{section8}. In appendix~\ref{section9}, we demonstrate
that the Majorana pfaffian defined by the overlap-Dirac operator in
$2+8n$ and $4+8n$ dimensions is a non-negative function of a
gauge-field configuration. The lattice spacing will be denoted by~$a$
throughout this paper.

\section{Majorana decomposition in even dimensional space}\label{section2}

\subsection{Majorana decomposition}\label{section2.1}

Here we consider a compatibility between the euclidean Majorana decomposition
and the charge conjugation property of the overlap-Dirac
operator~(\ref{bxfive}) in even $d$~dimensional lattice gauge theory. The
overlap-Dirac operator satisfies the gamma-hermiticity
\begin{equation}
   D^\dagger=\gamma D\gamma\,,
\label{twoxone}
\end{equation}
and the Ginsparg-Wilson relation
\begin{equation}
   \gamma D+D\gamma=aD\gamma D\,.
\label{twoxtwo}
\end{equation}
The operator is also local for ``admissible'' gauge-field
configurations~\cite{Hernandez:1999et}.

\pagebreak[3]

What we mean by the Majorana decomposition is a decomposition of the
action\footnote{In this paper, we consider only the kinetic term of
  fermions.  If other terms such as Yukawa couplings exist, the chiral
  symmetry for example may give rise to further
  restrictions~\cite{Fujikawa:2001ka}.} of the \emph{Dirac} fermion
\begin{equation}
   S_{\rm F}=a^d\sum_x\overline\psi(x)D\psi(x)\,,
\label{twoxthree}
\end{equation}
into two independent systems. Imitating the continuum
case~(\ref{onextwo}), we first set
\begin{equation}
   \psi={1\over\sqrt2}(\chi+i\eta),\qquad
   \overline\psi={1\over\sqrt2}(\chi^TB-i\eta^TB)\,,
\label{twoxfour}
\end{equation}
where $B$ denotes either $B_1$ or~$B_2$. For the overlap-Dirac
operator~$D$, from eq.~(\ref{bxsix}), one finds\footnote{Throughout
  this paper, the transpose and the conjugate operations on an
  operator are defined with respect to the corresponding kernel in
  position space.}
\begin{eqnarray}
   &&(B_2D)^T=+B_2D\,,\qquad\hbox{for $d=8n$}\,,
\label{twoxfive}\\
   &&(B_1D)^T=-B_1D\,,\qquad\hbox{for $d=2+8n$}\,,
\label{twoxsix}\\
   &&(B_2D)^T=-B_2D\,,\qquad\hbox{for $d=4+8n$}\,.
\label{twoxseven}
\end{eqnarray}
Therefore, for $d=2+8n$ and for $d=4+8n$, the action is decomposed as
\begin{equation}
   S_{\rm F}=a^d\sum_x\left[{1\over2}\chi^T(x)BD\chi(x)
     +{1\over2}\eta^T(x)BD\eta(x)\right],
\label{twoxeight}
\end{equation}
by using $B=B_1$ and $B=B_2$, respectively. However, for $d=8n$, the
above simple prescription for the Majorana decomposition does not
work. We recall that the Majorana decomposition worked in
unregularized continuum euclidean theory also for~$d=8n$, because
$(B_1\gamma_\mu D_\mu)^T=-B_1\gamma_\mu D_\mu$ holds; one can use
$B_1$ instead of~$B_2$ in these dimensions. For the overlap-Dirac
operator, on the other hand, $(B_1D)^T\neq-B_1D$ and this is
impossible.

For $d=2+8n$ and for $d=4+8n$, quantum theory of the Majorana fermion
can thus be defined by the functional integral
\begin{equation}
   \langle{\cal O}\rangle_{\rm M}=\int\prod_x\rmd\chi(x)\,{\cal O}
   \exp\left[-a^d\sum_x\,{1\over2}\chi^T(x)BD\chi(x)\right],
\label{twoxnine}
\end{equation}
in particular, the partition function is given by
\begin{equation}
   \langle1\rangle_{\rm M}=\int\prod_x\rmd\chi(x)\,
   \exp\left[-a^d\sum_x\,{1\over2}\chi^T(x)BD\chi(x)\right] =\Pf(BD)\,.
\label{twoxten}
\end{equation}
We emphasize that this construction is ``perfect'' in a sense that it
is given by the functional integral of a local action and the fermion
partition function is a smooth (single-valued) gauge invariant
function of a gauge-field configuration. In appendix~\ref{section9},
we show that the pfaffian~(\ref{twoxten}) is moreover
non-negative. Since the above provides a perfect definition of the
theory, one can conclude that Majorana fermions in $2+8n$ and~$4+8n$
dimensions are free from any pathologies in quantum theory, such as
the global gauge anomaly. A special case $d=4$ has been studied
in~ref.~\cite{Suzuki:2000ku}.  See
also~refs.~\cite{Neuberger:1997bg,Kaplan:1999jn}. We note that a
similar analysis as above for the Wilson-Dirac operator~(\ref{bxtwo})
leads to an identical conclusion; $d=2+8n$ and~$d=4+8n$ cases are
possible (besides a potential problem of a positivity of the pfaffian)
but $d=8n$ are not. In fact, in the context of supersymmetric
theories, a number of numerical simulations for Majorana fermions in 2
and~4 dimensions have been performed by using the Wilson-Dirac
operator~\cite{Feo:2002yi}.

\subsection{Impossibility of the Majorana decomposition for $d=8n$}\label{section2.2}

In the preceding subsection, we found that the simple Majorana
decomposition~(\ref{twoxfour}) does not work for~$d=8n$. Actually,
under some assumptions, it can be shown that there is no possible
``lattice modification'' of~eq.~(\ref{twoxfour}) which leads to a
successful decomposition. To show this, we set
\begin{equation}
   \psi={1\over\sqrt2}(\chi+i\eta),\qquad
   \overline\psi={1\over\sqrt2}(\chi^TB_1X-i\eta^TB_1X)\,,
\label{twoxeleven}
\end{equation}
where $X$ represents a possible modification of the charge conjugation
matrix in lattice theory, $X=1+O(a)$. We allow $X$ to depend on the
gauge field through the overlap-Dirac operator
\begin{equation}
   X=x(aD)\,,
\label{twoxtwelve}
\end{equation}
where a regular function~$x$ takes bounded values for all eigenvalues
of the operator~$aD$. For a classification of the function~$x$, it is
convenient to introduce operators
\begin{equation}
   H=\gamma D\,,
\label{twoxthirteen}
\end{equation}
and
\begin{equation}
   \Gammait=\gamma\left(1-{a\over2}D\right)=\gamma-{a\over2}H\,.
\label{twoxfourteen}
\end{equation}
(These $H$ and~$\Gammait$ are hermitean, because of the
gamma-hermiticity~(\ref{twoxone}).) An important property
of~$\Gammait$ which follows from the Ginsparg-Wilson
relation~(\ref{twoxtwo}) is
\begin{equation}
   \Gammait H+H\Gammait=0\,.
\label{twoxfifteen}
\end{equation}
We also define
\begin{equation}
   \tilde X=\gamma X\gamma\,.
\label{twoxsixteen}
\end{equation}
Then by noting relations, $\gamma=\Gammait+aH/2$, $aD=\Gammait
aH+a^2H^2/2$ and~$\Gammait^2=1-a^2H^2/4$, the above
assumption~(\ref{twoxtwelve}) can equivalently be expressed as
\begin{equation}
   \tilde X=h(a^2H^2)+\Gammait aHk(a^2H^2)\,,
\label{twoxseventeen}
\end{equation}
where regular functions~$h$ and~$k$ take bounded values for all
eigenvalues of the operator~$a^2H^2$.

Now, for the Majorana decomposition~(\ref{twoxeleven}) to work, we
have to have~$(B_1XD)^T=-B_1XD$ or
\begin{equation}
   (B_2\tilde XH)^T=-B_2\tilde XH\,.
\label{twoxeighteen}
\end{equation}
Then by substituting eq.~(\ref{twoxseventeen}) into this expression
and by using relations, $H^T=B_1HB_1^{-1}$ and~$\Gammait^T=B_1\Gammait
B_1^{-1}$ and~eq.~(\ref{twoxfifteen}), it is easy to see that
\begin{equation}
   a^2H^2[h(a^2H^2)+2\Gammait^2 k(a^2H^2)]=0\,,
\label{twoxnineteen}
\end{equation}
and thus
\begin{equation}
   \tilde X=-2\gamma\Gammait k(a^2H^2)\,.
\label{twoxtwenty}
\end{equation}
At this point, we note that the operator~$\Gammait$ always has a mode
such that $\Gammait\varphi=0$ (and thus $aH\varphi=\pm2\varphi$) in
the free theory with an infinite lattice volume. For the overlap-Dirac
operator, this may be seen from the explicit form~(\ref{bxfive}). This
property of~$\Gammait$ can also be shown on more general grounds by
repeating the argument of ref.~\cite{Fujikawa:2002is} which invokes
the Nielsen--Ninomiya theorem~\cite{Nielsen:1980rz}. This shows that
the kinetic term of the Majorana field~$\chi$,
\begin{equation}
   \chi^TB_1XD\chi=\chi^TB_2\tilde XH\chi =-2\chi^TB_2\gamma\Gammait
   k(a^2H^2)H\chi\,,
\label{twoxtwentyone}
\end{equation}
vanishes for $\chi=\varphi$. This mode~$\varphi$ is thus nothing but
the species doubler. We conclude that there is no possible
modification~$X$ which leads to a successful decomposition while being
free from the species doubling.

Under some assumptions, we have observed that the Majorana
decomposition does not work in $8n$~dimensional lattice gauge
theories. As a possible option, we may define the partition function
of the Majorana fermion by the square-root of the Dirac determinant
\begin{equation}
   \langle1\rangle_{\rm M}=\sqrt{\det D}\,.
\label{twoxtwentytwo}
\end{equation}
However, this prescription is potentially dangerous in two
aspects. First, the locality of the definition is not obvious. Namely,
it is not clear if $\sqrt{\det D}$ can be expressed as the functional
integral of a certain local (in a sense of lattice theory) action. For
a recent analysis of a similar problem in the square-root prescription
for the staggered fermion, see~ref.~\cite{Bunk:2004br}.

Another potential problem is a smoothness (and a single-valued-ness)
of the partition function as a function of a gauge-field
configuration. Although $\det D$ is a gauge invariant real number, it
is not obvious if the square-root can be defined smoothly everywhere
in the configuration space of lattice gauge fields. There may occur a
certain discrepancy depending on a path in the configuration space
along which one defines the square-root. This phenomenon, if it
occurs, is analogous to the global gauge
anomaly~\cite{Witten:fp,Elitzur:1984kr}. In fact, there is a good
reason to believe an occurrence of this phenomenon and we will give an
argument for this in a later section.

\section{Majorana-Weyl decomposition in $2+8n$ dimensions}\label{section3}

As we have seen in the introduction, in unregularized continuum theory
in~$2+8n$ dimensions, one can apply the Majorana decomposition on the
Weyl fermion to define a system of Majorana-Weyl fermions. On the
lattice, it is not trivial to define even Weyl fermions in the
classical level and this has becomes possible only with recent
developments~\cite{Kaplan:1992bt,Narayanan:1993wx,Narayanan:1998uu},
\cite{Luscher:1999du}--\cite{Kikukawa:2001mw}. First we briefly
summarize on Weyl fermions defined by using a lattice Dirac operator
which satisfies the Ginsparg--Wilson relation. Then, by using the
overlap-Dirac operator, we show that a further reduction or
decomposition of the Weyl fermion into Majorana-type degrees of
freedom exhibits a difficulty.

\subsection{Definition of Weyl fermions in lattice gauge theory}\label{section3.1}

In lattice gauge theories, it is possible to impose the Weyl chirality
condition in a consistent way, at least in the classical level, if a
Dirac operator~$D$ satisfies the Ginsparg--Wilson
relation~(\ref{twoxtwo}) and the gamma-hermiticity~(\ref{twoxone}). If
this is the case, one can introduce a modified chiral matrix
by~\cite{Narayanan:1998uu}
\begin{equation}
   \hat\gamma=\gamma(1-aD)\,,\qquad \hat\gamma^2=1\,,\qquad
   \hat\gamma^\dagger=\hat\gamma\,,
\label{threexone}
\end{equation}
where the last two relations hold due to eqs.~(\ref{twoxtwo})
and~(\ref{twoxone}). The chirality projection operators are then
defined by
\begin{equation}
   \hat P_\pm={1\pm\hat\gamma\over2},\qquad P_\pm={1\pm\gamma\over2}\,.
\label{threextwo}
\end{equation}
A crucial property of~$\hat P_{\pm}$ (which is again a result of the
Ginsparg--Wilson relation) is
\begin{equation}
   D\hat P_\mp=P_\pm D\,,
\label{threexthree}
\end{equation}
and one can consistently impose the chirality constraint
\begin{equation}
   \hat P_-\psi=\psi,\qquad\overline\psi P_+=\overline\psi\,,
\label{threexfour}
\end{equation}
in the action of the Dirac fermion
\begin{equation}
   S_{\rm F}=a^d\sum_x\overline\psi(x)D\psi(x)\,.
\label{threexfive}
\end{equation}
This defines a single left-handed Weyl fermion in the classical
level. We emphasize again that this consistent definition of Weyl
degrees of freedom in a lattice action has become possible only after
recent developments. For example, the Wilson-Dirac operator does not
allow such a construction.

\subsection{Difficulty of the Majorana-Weyl decomposition}\label{section3.2}

Let us examine a validity of the Majorana decomposition for Weyl
fermions by using the overlap-Dirac operator~(\ref{bxfive}). In
$2+8n$~dimensions, one has
\begin{equation}
   \hat\gamma^T=-B_2\hat\gamma B_2^{-1}\,,\qquad
   \hat P_\pm^T=B_2\hat P_\mp B_2^{-1}\,.
\label{threexsix}
\end{equation}
Now, by imitating the continuum case~(\ref{onextwo}), we may first try
\begin{equation}
   \psi={1\over\sqrt2}(\chi+i\eta),\qquad
   \overline\psi={1\over\sqrt2}(\chi^TB-i\eta^TB)\,,
\label{threexseven}
\end{equation}
where $B$ denotes either $B_1$ or~$B_2$. But this definition
immediately leads to a trouble because it does not provide a
consistent chirality for~$\chi$ and for~$\eta$; projection
operators~$\hat P_-$ and~$P_+$ are not exchanged under an action
of~$B$. $\hat P_-\chi=\chi$, for example, does not imply
$(\chi^TB)P_+=(\chi^TB)$.

The above definition of the modified chiral matrix~(\ref{threexone})
is not unique. One may consider, for example, a one-parameter family
of modified~$\gamma$ matrices
\begin{equation}
   \gamma^{(t)}={\gamma(1-taD)\over\sqrt{1-t(1-t)a^2D^\dagger D}},\qquad
   \overline\gamma^{(t)}=\gamma\gamma^{(1-t)}\gamma\,.
\label{threexeight}
\end{equation}
Then relations $(\gamma^{(t)})^2=(\overline\gamma^{(t)})^2=1$
and~$D\gamma^{(t)}=-\overline\gamma^{(t)}D$ hold. By using this kind
of freedom in a definition of chirality projection operators~$\hat
P_-$ and~$P_+$, it could be possible to avoid the above
difficulty. Also the Majorana decomposition could be modified by a
``lattice way'' as~eq.~(\ref{twoxeleven}).  To examine these
possibilities, we set
\begin{equation}
   \hat P_-={1-V\over2}\,,\qquad
   \bar P_+={1+U\over2}\,,\qquad
   V^2=U^2=1\,,
\label{threexnine}
\end{equation}
where operators $V$ and $U$ are hermitean and we impose
\begin{equation}
   \hat P_-\psi=\psi\,,\qquad\overline\psi\bar P_+=\overline\psi\,,
\label{threexten}
\end{equation}
instead of eq.~(\ref{threexfour}). Also we generalize the Majorana
decomposition as\footnote{In this expression, we may start with $B_2$
  instead of~$B_1$. The following analysis can be repeated for this
  case also and a similar conclusion is resulted.}
\begin{equation}
   \psi={1\over\sqrt2}(\chi+i\eta)\,,\qquad
   \overline\psi={1\over\sqrt2}(\chi^TB_1X-i\eta^TB_1X)\,,
\label{threexeleven}
\end{equation}
where $X=1+O(a)$. We assume that $U$ and~$X$ have following structures
\begin{equation}
   U=\gamma u(aD)\,,\qquad X=x(aD)\,,
\label{threextwelve}
\end{equation}
where regular functions $u$ and~$x$ take bounded values for all
eigenvalues of the operator~$aD$. We also define
\begin{equation}
   \tilde U=\gamma U\gamma\,,\qquad\tilde U^2=1\,,\qquad\tilde X=\gamma
   X\gamma\,.
\label{threexthirteen}
\end{equation}
Then by using operators $H$ and~$\Gammait$ in
eqs.~(\ref{twoxthirteen}) and~(\ref{twoxfourteen}), above
assumptions~(\ref{threextwelve}) can equivalently be expressed as
\begin{equation}
   \tilde U=aHf(a^2H^2)+\Gammait g(a^2H^2)\,,\qquad \tilde
   X=h(a^2H^2)+\Gammait aHk(a^2H^2)\,,
\label{threexfourteen}
\end{equation}
where regular functions $f$, $g$, $h$ and $k$ take bounded values for
all eigenvalues of the operator~$a^2H^2$.

Now, for the Weyl projection~(\ref{threexten}) to be consistent in the
action, we have to have $D\hat P_-=\bar P_+D$. This implies
\begin{equation}
   -HV=\tilde UH\,.
\label{threexfifteen}
\end{equation}
By substituting eq.~(\ref{threexfourteen}) into this, we find
\begin{equation}
   V=-aHf(a^2H^2)+\Gammait g(a^2H^2)\,,
\label{threexsixteen}
\end{equation}
by recalling eq.~(\ref{twoxfifteen}). Also, to ensure a consistent
chirality for the Majorana field, $\hat
P_-\chi=\chi\Rightarrow(\chi^TB_1X)\bar P_+= (\chi^TB_1X)$, we must
have (note that $B_1=\gamma B_2=-B_2\gamma$)
\begin{equation}
   -B_2^{-1}V^TB_2\tilde X=\tilde X\tilde U\,.
\label{threexseventeen}
\end{equation}
This implies
\begin{equation}
   aH[h(a^2H^2)f(a^2H^2)-\Gammait^2k(a^2H^2)g(a^2H^2)]=0\,,
\label{threexeighteen}
\end{equation}
by noting $H^T=-B_2HB_2^{-1}$ and~$\Gammait^T=-B_2\Gammait B_2^{-1}$
in $2+8n$~dimensions. Note that this relation shows
\begin{equation}
   \tilde X\tilde U
   =[h(a^2H^2)g(a^2H^2)+a^2H^2k(a^2H^2)f(a^2H^2)]\Gammait\,.
\label{threexnineteen}
\end{equation}
Finally, for the Majorana decomposition to work, we must have
$(B_1XD)^T=-B_1XD$ and
\begin{equation}
   HB_2^{-1}\tilde X^TB_2=\tilde XH\,,
\label{threextwenty}
\end{equation}
but this requirement sets no restriction on the functions $h$ and~$k$.

As noted in section~\ref{section2.2}, the operator~$\Gammait$ has a
mode such that $\Gammait\varphi=0$ and $aH\varphi=\pm2\varphi$. Above
relations then imply
\begin{equation}
   \tilde U\varphi=\pm2f(4)\varphi\,,\qquad \tilde
   X\varphi=h(4)\varphi\,,\qquad f(4)h(4)=0\,,
\label{threextwentyone}
\end{equation}
where the last relation follows from eq.~(\ref{threexnineteen}). From
these expressions, a trouble becomes clear: If $f(4)=0$, then we
cannot set $\tilde U^2=V^2=1$ on~$\varphi$. Namely, the projection
operator cannot be properly defined. On the other hand, if $h(4)=0$,
the kinetic term of the Majorana field,
$\chi^TB_1XD\chi=-\chi^TB_2\tilde XH\chi$, vanishes for
$\chi=\varphi$. This mode $\varphi$ is nothing but the species
doubler. An interesting example in the latter case is the choice,
$U=\gamma$, $V=\hat\gamma$ and~$X=\gamma\Gammait$. It can be seen that
this choice fulfills all the above requirements for a consistent
Majorana decomposition, but the kinetic term acquires doubler's zeros.

Summarizing our analysis so far, we have observed that there is always
a conflict between the Weyl projection of the type~(\ref{threexten})
based on the Ginsparg-Wilson relation and the modified Majorana
decomposition~(\ref{threexeleven}) in~$2+8n$ dimensions.

Going back to the standard choice~(\ref{threexone}), to define the
partition function of the Weyl fermion
\begin{equation}
   \langle1\rangle_{\rm W}
   =\int\rmD[\psi]\rmD[\overline\psi]\,e^{-S_{\rm F}}\,,\qquad S_{\rm
     F}=a^d\sum_x\overline\psi(x)D\psi(x)\,,
\label{threextwentytwo}
\end{equation}
one introduces an orthonormal complete set of vectors in the
constrained space~(\ref{threexfour}),
\begin{eqnarray}
   &&\hat P_-v_j(x)=v_j(x)\,,\qquad
   (v_k,v_j)=a^d\sum_xv_k(x)^\dagger v_j(x)=\delta_{kj}\,,
\label{threextwentythree}\\
   &&\overline v_k(x)P_+=\overline v_k(x)\,,\qquad
   (\overline v_k^\dagger,\overline v_j^\dagger)=\delta_{kj}\,.
\label{threextwentyfour}
\end{eqnarray}
We note
\begin{equation}
   \Tr\hat\gamma=\Tr\hat\gamma^T=-\Tr B_2\hat\gamma B_2^{-1}
   =-\Tr\hat\gamma=0\,,
\label{threextwentyfive}
\end{equation}
due to the reality of the representation, and $\Tr\hat P_-=\Tr
P_+=(1/2)\Tr1$.  Thus indices~$j$, $k$ in
eqs.~(\ref{threextwentythree}) and~(\ref{threextwentyfour}) run over
from~1 to~$(1/2)\Tr1$ that is a fixed number being independent of a
gauge-field configuration.\footnote{Since $\Tr1=\sum_x\tr1$
  and~$\tr1=2^{d/2}$, $\Tr1$ is always an even positive integer.} By
expanding fermion fields in terms of these bases
\begin{equation}
   \psi(x)=\sum_jv_j(x)c_j\,,\qquad \overline\psi(x)=\sum_k\overline
   c_k\overline v_k(x)\,,
\label{threextwentysix}
\end{equation}
the integration measure is defined by
\begin{equation}
   \rmD[\psi]\rmD[\overline\psi]=\prod_j\rmd c_j\prod_k\rmd\overline
   c_k\,.
\label{threextwentyseven}
\end{equation}
Then the partition function~(\ref{threextwentytwo}) is given by
\begin{equation}
   \langle1\rangle_{\rm W}=\det M,\qquad M_{kj} =a^d\sum_x\overline
   v_k(x)Dv_j(x)={2\over a}a^d\sum_x\overline v_k(x)v_j(x)\,.
\label{threextwentyeight}
\end{equation}
In deriving the last expression, we have noted the relation~$P_+D\hat
P_-=(2/a)P_+\hat P_-$.

We have observed that the Majorana decomposition of Weyl fermions has
a difficulty. As a possible option, we may define the partition
function of Majorana-Weyl fermions by the square-root of the Weyl
determinant
\begin{equation}
   \langle1\rangle_{\rm MW}=\sqrt{\det M}\,.
\label{threextwentynine}
\end{equation}
However, as the square-root prescription for Majorana
fermions~(\ref{twoxtwentytwo}), this prescription is potentially
dangerous; the locality and the smoothness are not obvious. An
analysis of smoothness in the present case is not easy, because $\det
M$ depends on a choice of the basis
vectors~$\{v_j\}$~(\ref{threextwentythree}). A system of a single
Majorana-Weyl fermion is always anomalous and one has to introduce
left- and right-handed Majorana-Weyl fermions to eliminate the local
(perturbative) gauge anomaly in continuum theory. Then one has to
examine if it is possible to choose basis vectors such that $\det M$
is gauge invariant. The choice must also preserve the locality. See
refs.~\cite{Luscher:1999du,Luscher:1999un} for a proper way to
formulate the problem. In a non-perturbative level, such an ``ideal''
basis has been constructed only for $\U(1)$ gauge
theories~\cite{Luscher:1999du}.\footnote{As a recent attempt towards
  an implementation of this construction in actual numerical
  simulations, see ref.~\cite{Kadoh:2003ii}. As a completely different
  kind of approach to lattice chiral gauge theories, see
  ref.~\cite{Golterman:2004qv}.} It might be possible to study the
present problem of smoothness by using real representations of
$\SO(2)$ gauge theories.

If the matrix~$M_{kj}$ in eq.~(\ref{threextwentyeight}) is
anti-symmetric then the square-root in eq.~(\ref{threextwentynine})
will~be
\begin{equation}
   \langle1\rangle_{\rm MW}=\Pf M\,,
\label{threexthirty}
\end{equation}
\looseness=1 and the above problem of smoothness is removed. Note again that,
however, the matrix~$M$ depends on a choice of basis
vectors~$\{v_j\}$. \pagebreak[3] For a generic choice of basis vectors, the
matrix~$M$ is neither anti-symmetric nor real. Again one first has to
seek an ideal basis and study its property before examining if this
factorization of the chiral determinant~$\det M$ occurs.

It is interesting to note that, in the overlap formalism for real Weyl
fermions in~$2+8n$ dimensions, it has been known that many-body
hamiltonians, from which lowest-energy states in the overlap formalism
are defined, are decomposed into two independent
systems~\cite{Huet:1996pw}. If this fact implies a simple
factorization (such as the pfaffian) of the Weyl determinant, it would
provide a way out from above difficulties. We will examine this
possibility in the next section; our conclusion will, however, be
negative.

\section{Majorana-Weyl fermions based on the overlap formalism}\label{section4}

\subsection{Brief overview of the overlap formalism}\label{section4.1}

First we briefly recall basics of the overlap formalism. We assume
that $d=2+8n$ and the gauge-group representation is real. In the
overlap formalism, the Weyl determinant is defined as an overlap of
two states $\langle L{-}|L{+}\rangle$, each of the states is the
ground state of many-body hamiltonians ${\cal H}^+$ and~${\cal H}^-$,
respectively. These ${\cal H}^\pm$ are defined by
\begin{equation}
   {\cal H}^\pm=a^d\sum_xa(x)^\dagger H^\pm a(x)\,,
\label{fourxone}
\end{equation}
where hermitean Wilson-Dirac operators~$H^\pm$ are given by
\begin{equation}
   H^+=\gamma\left({1\over a}-D_{\rm W}\right),\qquad
   H^-=\gamma(-m-D_{\rm W})\,.
\label{fourxtwo}
\end{equation}
The mass parameter~$m$ is taken to be infinity~$m\to+\infty$ in the
final expressions of the overlap. The creation and annihilation
operators obey the anti-commutation relations
\begin{equation}
   \{a(x),a(y)^\dagger\}=a^{-d}\delta_{x,y}\,,\qquad
   \{a(x),a(y)\}=0\,.
\label{fourxthree}
\end{equation}
To find the lowest-energy states~$|L\pm\rangle$, we define
eigenfunctions of~$H^\pm$ with negative eigenvalues:
\begin{equation}
   H^\pm v_j^\pm(x)=-\lambda_j^\pm
   v_j^\pm(x)\,,\qquad\lambda_j^\pm>0\,,\qquad
   (v_k^\pm,v_j^\pm)=a^d\sum_x v_k^\pm(x)^\dagger
   v_j^\pm(x)=\delta_{kj}\,.
\label{fourxfour}
\end{equation}
Due to the charge conjugation property of the hermitean Wilson-Dirac
operators in $2+8n$~dimensions (see eq.~(\ref{bxfour})), we have
\begin{equation}
   H^\pm B_2^{-1}v_j^\pm(x)^*=+\lambda_j^\pm B_2^{-1}v_j^\pm(x)^*\,.
\label{fourxfive}
\end{equation}
Therefore, eigenvalues of $H^+$, for example, always appear in pair as
$+\lambda_j^+$ and~$-\lambda_j^+$. We assume that $H^\pm$ have no zero
modes; otherwise lowest-energy states are not uniquely defined. Since
$H^\pm$ are hermitean, a set of vectors~$\{v_j^+,B_2^{-1}v_j^{+*}\}$
or~$\{v_j^-,B_2^{-1}v_j^{-*}\}$ spans a complete set. These facts show
that, in eq.~(\ref{fourxfour}), the \pagebreak[3] index~$j$ runs from 1
to~$(1/2)\Tr1$. Expanding the annihilation operators by using each of
complete sets,
\begin{equation}
   a(x)=\sum_{j=1}^{\Tr(1/2)}[ a_j^\pm v_j^\pm(x)+{\tilde a}_j^\pm
     B_2^{-1}v_j^\pm(x)^*]\,,
\label{fourxsix}
\end{equation}
we have
\begin{equation}
   \{a_j^\pm,a_k^{\pm\dagger}\} =\{{\tilde a}_j^\pm,{\tilde
     a}_k^{\pm\dagger}\}=\delta_{jk}\,,\qquad
   \{a_j^\pm,a_k^\pm\}=\{{\tilde a}_j^\pm,{\tilde a}_k^\pm\}
   =\{a_j^\pm,{\tilde a}_k^\pm\} =\{a_j^\pm,{\tilde
     a}_k^{\pm\dagger}\}=0\,,
\label{fourxseven}
\end{equation}
and
\begin{equation}
   {\cal H}^\pm=\sum_{j=1}^{\Tr(1/2)}(
   -\lambda_j^\pm a_j^{\pm\dagger}a_j^\pm
   +\lambda_j^\pm{\tilde a}_j^{\pm\dagger}{\tilde a}_j^\pm)\,.
\label{fourxeight}
\end{equation}
In deriving above expressions, we have noted the fact that $v_j^+$
and~$B_2^{-1}v_k^{+*}$ for example are orthogonal, because they give
rise to different eigenvalues of the hermitean operator~$H^+$. From
this expression of hamiltonians, the lowest-energy states are given by
\begin{equation}
   |L{+}\rangle=a_1^{+\dagger}a_2^{+\dagger}\cdots
   a_{\Tr(1/2)}^{+\dagger}|0\rangle\,,\qquad
   |L{-}\rangle=a_1^{-\dagger}a_2^{-\dagger}\cdots
   a_{\Tr(1/2)}^{-\dagger}|0\rangle\,.
\label{fourxnine}
\end{equation}
Here it is important to note the fact that the Fock vacuum~$|0\rangle$
is common to~$|L{+}\rangle$ and to~$|L{-}\rangle$ because it is
specified by the condition~$a(x)|0\rangle=0$. The overlap of these two
states is thus given by
\begin{eqnarray}
   \langle L{-}|L{+}\rangle&=&
   \langle0|a_{\Tr(1/2)}^-\cdots a_2^-a_1^-
   a_1^{+\dagger}a_2^{+\dagger}\cdots
   a_{\Tr(1/2)}^{+\dagger}|0\rangle
\nonumber\\
   &=&\det m\,,
\label{fourxten}
\end{eqnarray}
where the matrix~$m$ is given by
\begin{equation}
   m_{kj}=a^d\sum_x v_k^-(x)^\dagger v_j^+(x)\,.
\label{fourxeleven}
\end{equation}

Comparing the overlap-Dirac operator~(\ref{bxfive}) and the hermitean
Wilson-Dirac operators~(\ref{fourxtwo}), we see that
$\hat\gamma=H^+/\sqrt{(H^+)^2}$
and~$\gamma=-\lim_{m\to+\infty}H^-/\sqrt{(H^-)^2}$. Therefore the
chirality constraints~(\ref{threextwentythree})
and~(\ref{threextwentyfour}) and the conditions for
eigenfunctions~(\ref{fourxfour}) turn out to be identical; we can thus
identify $v_j=v_j^+$ and~$\overline v_k^\dagger=v_k^-$. With this
identification, $M_{kj}=(2/a)m_{kj}$ and the chiral determinants given
by eq.~(\ref{threextwentyeight}) and by eq.~(\ref{fourxten}) are
identical, up to a proportionality constant.

\subsection{Overlap formulation of Majorana-Weyl fermions}\label{section4.2}

Now, for $2+8n$~dimensions, the many-body hamiltonians~${\cal H}^\pm$
enjoy a decomposition into two mutually independent
systems~\cite{Huet:1996pw}; this is analogous to the Majorana
decomposition. To see this, we define hermitean parts of the creation
and annihilation operators by
\begin{equation}
   \gamma(x)={1\over\sqrt{2}}
   \left[B_2^{1/2}a(x)+B_2^{-1/2}a(x)^\dagger\right],\qquad
   \delta(x)={1\over\sqrt{2}i}
   \left[B_2^{1/2}a(x)-B_2^{-1/2}a(x)^\dagger\right].
\label{fourxtwelve}
\end{equation}
Here we have introduced matrices $B_2^{1/2}$ and~$B_2^{-1/2}$ as
objects such that $(B_2^{1/2})^2=B_2$, $(B_2^{-1/2})^2=B_2^{-1}$
and~$B_2^{1/2}B_2^{-1/2}=1$. Such matrices are not unique but we can
adopt
\begin{equation}
   B_2^{1/2}={1\over\sqrt{2}}(1+B_2)\,,\qquad
   B_2^{-1/2}={1\over\sqrt{2}}(1-B_2)\,,
\label{fourxthirteen}
\end{equation}
because $B_2^2=-1$ holds in our representation (see
eq.~(\ref{axfourteen})).  Note that these $B_2^{1/2}$ and~$B_2^{-1/2}$
are symmetric matrices. It is then easy to see that each of
$\gamma(x)$ and~$\delta(x)$ generates the Clifford algebra including
space indices:
\begin{equation}
   \{\gamma(x),\gamma(y)\}=\{\delta(x),\delta(y)\}=a^{-d}\delta_{x,y}\,,\qquad
   \{\gamma(x),\delta(y)\}=0\,.
\label{fourxfourteen}
\end{equation}
Also, by defining
\begin{equation}
   H_{\rm MW}^\pm=B_2^{1/2}H^\pm B_2^{-1/2}\,,
\label{fourxfifteen}
\end{equation}
and by noting $(H_{\rm MW}^\pm)^T=-H_{\rm MW}^\pm$, we have the
decomposition
\begin{equation}
   {\cal H}^\pm={1\over2}a^d\sum_x\gamma(x)H_{\rm MW}^\pm\gamma(x)
   +{1\over2}a^d\sum_x\delta(x)H_{\rm MW}^\pm\delta(x)\,.
\label{fourxsixteen}
\end{equation}
Thus the many-body hamiltonians are decomposed into two mutually
independent systems. One would then expect that this decomposition,
after applying the overlap formula only to (say) a system
of~$\gamma(x)$, leads to a natural factorization of the overlap
formula~(\ref{fourxten}) (such as the pfaffian) for Weyl
fermions. However, the real situation is more subtle as we will see
below.

To see what happens clearly, we further expand~$\gamma(x)$ by using
eigenfunctions~(\ref{fourxfour}),
\begin{eqnarray}
   \gamma(x)&=&B_2^{1/2}\sum_{j=1}^{\Tr(1/2)}
   \left[\xi_j^+ v_j^+(x)+\xi_j^{+\dagger}B_2^{-1}v_j^+(x)^*\right]
\nonumber\\
   &=&B_2^{1/2}\sum_{j=1}^{\Tr(1/2)}
   \left[\xi_j^- v_j^-(x)+\xi_j^{-\dagger}B_2^{-1}v_j^-(x)^*\right].
\label{fourxseventeen}
\end{eqnarray}
Then we have
\begin{equation}
   \{\xi_j^+,\xi_k^{+\dagger}\}=\delta_{j,k}\,,\qquad
   \{\xi_j^+,\xi_k^+\}=0\,,\qquad
   \{\xi_j^-,\xi_k^{-\dagger}\}=\delta_{j,k}\,,\qquad
   \{\xi_j^-,\xi_k^-\}=0\,,
\label{fourxeighteen}
\end{equation}
and
\begin{eqnarray}
   {\cal H}_{\rm MW}^+&=&{1\over2}a^d\sum_x\gamma(x)H_{\rm MW}^+\gamma(x)
   =\sum_{j=1}^{\Tr(1/2)}(-\lambda_j^+)
   \left(\xi_j^{+\dagger}\xi_j^+-{1\over2}\right),
\label{fourxnineteen}\\
   {\cal H}_{\rm MW}^-&=&{1\over2}a^d\sum_x\gamma(x)H_{\rm MW}^-\gamma(x)
   =\sum_{j=1}^{\Tr(1/2)}(-\lambda_j^-)
   \left(\xi_j^{-\dagger}\xi_j^--{1\over2}\right).
\label{fourxtwenty}
\end{eqnarray}
From these expressions, the lowest-energy states of these
``Majorana-Weyl'' hamiltonians are given by
\begin{eqnarray}
   |L{+}\rangle_{\rm MW}
   &=&\xi_1^{+\dagger}\xi_2^{+\dagger}\cdots\xi_{\Tr(1/2)}^{+\dagger}
   |0\rangle_+,
\label{fourxtwentyone}\\
   |L{-}\rangle_{\rm MW}
   &=&\xi_1^{-\dagger}\xi_2^{-\dagger}\cdots\xi_{\Tr(1/2)}^{-\dagger}
   |0\rangle_-\,.
\label{fourxtwentytwo}
\end{eqnarray}
The important point to recognize here is that the Fock vacua,
$|0\rangle_+$ and~$|0\rangle_-$, are generally different; the former
is defined by the condition~$\xi_j^+|0\rangle_+=0$ and the latter is
defined by~$\xi_j^-|0\rangle_-=0$. Since $\xi_j^+$ and~$\xi_j^-$ are
related by a non-trivial Bogoliubov transformation as we will see
shortly, a relation between $|0\rangle_+$ and~$|0\rangle_-$ is also
non-trivial. This is a crucial difference from the overlap formula for
Weyl fermions. Due to this fact, there emerge non-polynomial
dependences of the Majorana-Weyl overlap~${}_{\rm MW}\langle
L{-}|L{+}\rangle_{\rm MW}$ on eigenfunctions~(\ref{fourxfour}).

The Bogoliubov transformation is given by
\begin{equation}
   \xi_k^-=\sum_{j=1}^{\Tr(1/2)}(m_{kj}\xi_j^++n_{kj}\xi_j^{+\dagger})\,,
\label{fourxtwentythree}
\end{equation}
where
\begin{equation}
   m_{kj}=a^d\sum_x v_k^-(x)^\dagger v_j^+(x)\,,\qquad
   n_{kj}=a^d\sum_x v_k^-(x)^\dagger B_2^{-1}v_j^+(x)^*\,.
\label{fourxtwentyfour}
\end{equation}
These matrices satisfy
\begin{eqnarray}
   mm^\dagger+nn^\dagger&=&1\,,
\label{fourxtwentyfive}\\
   mn^T+nm^T&=&0\,,
\label{fourxtwentysix}
\end{eqnarray}
to ensure the identical anti-commutation
relations~(\ref{fourxeighteen}) for $\xi_j^+$ and for~$\xi_j^-$. The
Bogoliubov transformation can equally be represented as
\begin{equation}
   \xi_k^-=e^{-{\cal G}}\sum_{j=1}^{\Tr(1/2)}m_{kj}\xi_j^+\,e^{{\cal
       G}}\,, \qquad {\cal G}=
      {1\over2}\sum_{j,k}\xi_j^{+\dagger}(m^{-1}n)_{jk}\xi_k^{+\dagger}\,,
\label{fourxtwentyseven}
\end{equation}
where this construction of ${\cal G}$ is consistent because the
matrix~$m^{-1}n$ is anti-symmetric
from~eq.~(\ref{fourxtwentysix}). Two Fock vacua are thus related as
\begin{equation}
   |0\rangle_-=Ce^{-{\cal G}}|0\rangle_+\,,
\label{fourxtwentyeight}
\end{equation}
and the normalization constant~$C$ is determined by\footnote{In
  deriving the last expression, we have used the formula which holds
  for fermionic creation and annihilation operators,
\begin{equation}
   \langle0|
   \exp\left({1\over2}a{\cal M}a\right)
   \exp\left({1\over2}a^\dagger{\cal N}a^\dagger\right)|0\rangle
   =\det\nolimits^{1/2}(1+{\cal M}{\cal N})\,.
\label{fourxtwentynine}
\end{equation}
This formula may readily be proven by going to the coherent state
representation.}
\begin{eqnarray}
   1&=&|C|^2{}_+\langle0|e^{-{\cal G}^\dagger}e^{-{\cal G}}|0\rangle_+
\nonumber\\
   &=&|C|^2
   {}_+\langle0|
   \exp\left[{1\over2}\sum_{j,k}\xi_j^+(m^{-1*}n^*)_{jk}\xi_k^+\right]
   \exp\left[-{1\over2}\sum_{l,m}
   \xi_l^{+\dagger}(m^{-1}n)_{lm}\xi_m^{+\dagger}\right]
   |0\rangle_+
\nonumber\\
   &=&|C|^2\det\nolimits^{1/2}(1-m^{-1*}n^*m^{-1}n)\,.
\label{fourxthirty}
\end{eqnarray}
We next note
\begin{eqnarray}
   \det(1-m^{-1*}n^*m^{-1}n)&=&\det(1+m^{-1*}n^*n^Tm^{-1T})
\nonumber\\
   &=&\det(1+m^{-1}nn^\dagger m^{-1\dagger})
\nonumber\\
   &=&\det[m^{-1}(mm^\dagger+nn^\dagger)m^{-1\dagger}]
\nonumber\\
   &=&|{\det m}|^{-2}\,,
\label{fourxthirtyone}
\end{eqnarray}
where the last equality follows from
eq.~(\ref{fourxtwentyfive}). Therefore we have
\begin{equation}
   C=|{\det m}|^{1/2}e^{i\theta}\,,
\label{fourxthirtytwo}
\end{equation}
where $\theta$ is an unspecified phase which may depend on a
gauge-field configuration. The Majorana-Weyl overlap thus gives rise
to
\begin{eqnarray}
   {}_{\rm MW}\langle L{-}|L{+}\rangle_{\rm MW}&=&
   {}_-\langle0|
   \xi_{\Tr(1/2)}^-\cdots\xi_2^-\xi_1^-
   \xi_1^{+\dagger}\xi_2^{+\dagger}\cdots\xi_{\Tr(1/2)}^{+\dagger}
   |0\rangle_+
\nonumber\\
   &=&C^*{}_+\langle0|e^{-{\cal G}^\dagger}e^{-{\cal G}}
   m_{\Tr(1/2)j}\xi_j^+\cdots m_{1k}\xi_k^+e^{{\cal G}}
   \xi_1^{+\dagger}\xi_2^{+\dagger}\cdots\xi_{\Tr(1/2)}^{+\dagger}
   |0\rangle_+
\nonumber\\
   &=&C^*\det m{}_+\langle0|e^{-{\cal G}^\dagger}e^{-{\cal G}}|0\rangle_+
\nonumber\\
   &=&C^{-1}\det m
\nonumber\\
   &=&|{\det m}|^{1/2}e^{i\vartheta},\qquad\vartheta=\arg\det m-\theta\,.
\label{fourxthirtythree}
\end{eqnarray}
We have described the calculation somewhat in detail to illustrate how
non-polynomial dependences on the eigenfunctions emerge. The final
result can be regarded as
\begin{equation}
   {}_{\rm MW}\langle L{-}|L{+}\rangle_{\rm MW}=\sqrt{\det m}\,,
\label{fourxthirtyfour}
\end{equation}
where a definition of the square-root amounts to specifying how the
phase~$\vartheta$ changes according to a change of gauge fields. This
result is certainly ensuring because a square of the Majorana-Weyl
overlap is the Weyl overlap. Nevertheless, the result is not
interesting because the Majorana-Weyl overlap eventually reduces to
the square-root prescription~(\ref{threextwentynine}). The difficulty
we faced in the previous section is not removed by the overlap
formalism.

\section{Odd dimensional cases}\label{section5}

\subsection{Majorana decomposition}\label{section5.1}

In this section, we consider the Majorana decomposition in odd
dimensional spaces. For $1+8n$~dimensions, as shown in
eqs.~(\ref{bxthree}) and~(\ref{bxsix}), both the Wilson-Dirac operator
and the overlap-Dirac operator have no simple transformation law under
the transpose operation.  Hence the Majorana decomposition cannot be
applied. Even we cannot study possible ``lattice modifications'' of
the charge conjugation matrix, due to lack of a definite charge
conjugation property of these lattice Dirac operators.

For $3+8n$~dimensions, the Wilson-Dirac operator has the desired
property
\begin{equation}
   (B_1D_{\rm W})^T=-B_1D_{\rm W}\,.
\label{fivexone}
\end{equation}
Therefore, the Majorana decomposition can be applied at least for the
Wilson-Dirac operator. Namely
\begin{equation}
   \langle1\rangle_{\rm M}=\int\prod_x\rmd\chi(x)\,
   \exp\left[-a^d\sum_x\,{1\over2}\chi^T(x)B_1D_{\rm W}\chi(x)\right]
   =\Pf(B_1D_{\rm W})\,.
\label{fivextwo}
\end{equation}
This immediately implies that a quantum theory of Majorana fermions in
$3+8n$~dimensions is free from any pathologies. Although this is a
``perfect'' definition, the pfaffian is neither positive nor real in
general. One way to see this is to recall that the Dirac
determinant~$\det D_{\rm W}$, being gauge invariant, suffers from the
parity anomaly~\cite{Redlich:1983kn,So:1984nf}.\footnote{We regard the
  parity anomaly as a property of the theory rather than a
  pathology. H.S. would like to thank Martin L\"uscher for an
  enlightening discussion on this point.} From the relation
\begin{equation}
   D_{\rm W}(x,y)^\dagger=D_{\rm W}(-x,-y)(U\to U^P)\,,
\end{equation}
where $U^P$ is the parity transformation of gauge fields,
\begin{equation}
   U^P(x,\mu)=U(-x-a\hat\mu,\mu)^{-1}\,,
\end{equation}
we have
\begin{equation}
   (\det D_{\rm W})^*=\det D_{\rm W}(U\to U^P)\,.
\end{equation}
The parity anomaly on the other hand implies
\begin{equation}
   \det D_{\rm W}(U^P)\neq\det D_{\rm W}(U)\,.
\end{equation}
Therefore, $\det D_{\rm W}$ is a complex number. Since the square of
the pfaffian~$\Pf(B_1D_{\rm W})$ is the Dirac determinant, the
pfaffian cannot be a real number in general.

\subsection{Comment on a prescription based on the overlap formalism}\label{section5.2}

In ref.~\cite{Maru:1997kh}, following the overlap formalism for Dirac
fermions in odd dimensions~\cite{Narayanan:1997by}, it was pointed out
that many-body hamiltonians in the overlap formalism enjoy a
decomposition when $d=3+8n$ (and when the gauge-group representation
is real). Here, we note that this decomposition does not imply a
factorized form of the Dirac determinant.  The overlap formalism for
the Dirac fermion in
$3+8n$~dimensions~\cite{Narayanan:1997by,Kikukawa:1997qh} starts with
hermitean Wilson-Dirac operators in a space with one dimension higher,
i.e., $H^\pm$ in $4+8n$~dimensions
\begin{equation}
   H^\pm=\gamma(\pm m-D_{\rm W})\,,
\label{fivexthree}
\end{equation}
but with a ``dimensional reduction'' to $3+8n$~dimensions. Here the
dimensional reduction means that the summation over~$\mu$ in the
Wilson-Dirac operator runs from $\mu=0$ to~$\mu=2+8n$ only:
\begin{equation}
   D_{\rm
     W}={1\over2}\sum_{\mu=0}^{2+8n}[\gamma_\mu(\nabla_\mu^*+\nabla_\mu)
     -a\nabla_\mu^*\nabla_\mu]\,.
\label{fivexfour}
\end{equation}
This operator does not contain gauge fields along the $\mu=3+8n$
direction although gamma matrices are those for $4+8n$~dimensions. The
many-body hamiltonians are then defined~by
\begin{equation}
   {\cal H}^\pm=a^d\sum_xa(x)^\dagger H^\pm a(x)\,,
\label{fivexfive}
\end{equation}
where the position~$x$ runs over a $3+8n$ (odd) dimensional
lattice. After applying the overlap prescription, this setup defines
the \emph{Dirac} fermion in $3+8n$~dimensions.

To see that above many-body hamiltonians are decomposed into two
independent systems, we define the matrix~$B_3$ (note that gamma
matrices are those of $4+8n$~dimensions) by
\begin{equation}
   B_3=B_1\gamma_{3+8n}\,,\qquad B_3^T=+B_3,\qquad B_3^2=-1\,.
\label{fivexsix}
\end{equation}
Due to the fact that $\mu=3+8n$ is omitted in the Wilson-Dirac
operator~(\ref{fivexfour}), one has
\begin{equation}
   (H^{\pm})^T=-B_3H^\pm B_3^{-1}\,.
\label{fivexseven}
\end{equation}
Therefore, by defining
\begin{equation}
   B_3^{1/2}={1\over\sqrt{2}}(1+B_3)\,,\qquad
   B_3^{-1/2}={1\over\sqrt{2}}(1-B_3)\,,
\label{fivexeight}
\end{equation}
and
\begin{equation}
   H_{\rm M}^{\pm}=B_3^{1/2}H^\pm B_3^{-1/2}\,,
\label{fivexnine}
\end{equation}
one has $(H_{\rm M}^{\pm})^T=-H_{\rm M}^{\pm}$. Finally, by setting
\begin{equation}
   \gamma(x)={1\over\sqrt{2}}
   [B_3^{1/2}a(x)+B_3^{-1/2}a(x)^\dagger]\,,\qquad
   \delta(x)={1\over\sqrt{2}i}
   [B_3^{1/2}a(x)-B_3^{-1/2}a(x)^\dagger]\,,
\label{fivexten}
\end{equation}
one has a decomposition of many-body hamiltonians
\begin{equation}
   {\cal H}^\pm={1\over2}a^d\sum_x\gamma(x)H_{\rm M}^\pm\gamma(x)
   +{1\over2}a^d\sum_x\delta(x)H_{\rm M}^\pm\delta(x)\,.
\label{fivexeleven}
\end{equation}
One clearly sees a similarity of expressions to those of
section~\ref{section4.2} with the replacement~$B_2\to B_3$. In
particular, since the operators~$\gamma(x)$ and~$\delta(x)$ obey
anti-commutation relations identical to eq.~(\ref{fourxfourteen}), we
have the same conclusion as section~\ref{section4.2}. An application
of the overlap prescription to a system of~$\gamma(x)$ reduces to the
square-root prescription; this time, it is the square-root of the
Dirac determinant defined by the overlap formalism.

For $1+8n$~dimensions, we found that the Majorana decomposition does
not work with either use of the Wilson-Dirac and the overlap-Dirac
operators. For these dimensions, we may try a use of the overlap
formalism. For $d=1+8n$, one starts with the hermitean Wilson
operator~$H^\pm=\gamma(\pm m-D_{\rm W})$ in $2+8n$ dimensions, but
with the dimensional reduction in the $\mu=1+8n$ direction:
\begin{equation}
   D_{\rm
     W}={1\over2}\sum_{\mu=0}^{8n}[\gamma_\mu(\nabla_\mu^*+\nabla_\mu)
     -a\nabla_\mu^*\nabla_\mu]\,.
\label{fivextwelve}
\end{equation}
Since the gamma matrices are those of $2+8n$~dimensions, all algebraic
relations in section~\ref{section4.2} hold as they stand. This is
expected because, after the dimensional reduction, Majorana-Weyl
fermions in $2+8n$~dimensions become Majorana fermions in
$1+8n$~dimensions. In particular, defining
\begin{equation}
   H_{\rm M}^\pm=B_2^{1/2}H^\pm B_2^{-1/2}\,,
\label{fivexthirteen}
\end{equation}
and using eq.~(\ref{fourxtwelve}), we have a decomposition
\begin{equation}
   {\cal H}^\pm={1\over2}a^d\sum_x\gamma(x)H_{\rm M}^\pm\gamma(x)
   +{1\over2}a^d\sum_x\delta(x)H_{\rm M}^\pm\delta(x)\,.
\label{fivexfourteen}
\end{equation}
Again an application of the overlap formalism to a system
of~$\gamma(x)$ will reduce to the square-root prescription and this
setup does not remove a difficulty of the Majorana decomposition in
$1+8n$~dimensions.

\section{Physical meaning of difficulties in $8n$, $1+8n$ and $2+8n$
dimensions}
\label{section6}

Let us go back to table~\ref{tab1} in section~\ref{section1} in which
results of our analysis are summarized. Entries with the
cross~$\times$ in this table are those for which we found
difficulties. In what follows, we give an argument that these
difficulties have an intrinsic physical meaning.

Let us start our discussion by recalling following facts. In
$d$~dimensional \emph{Minkowski} spacetime, when $d=0$, $4\bmod8$, the
Majorana fermion and the Weyl fermion in a \emph{real} representation
are equivalent. This means that one can convert the Majorana fermion
into the real Weyl fermion by multiplying the projection
operator~$(1-\gamma)/2$. This operation is invertible because one can
reproduce the original Majorana fermion by further multiplying
$1+{\cal C}$, where ${\cal C}$ is the charge conjugation operation.
For this, relations~${\cal C}^2=1$ and~$\{{\cal C},\gamma\}=0$ (the
charge conjugation flips the chirality) holding in these dimensions
are crucial.  Similarly, one can convert the real Weyl fermion into
the Majorana fermion by multiplying~$1+{\cal C}$; this operation is
again invertible. As expected from this equivalence, Weyl fermions in
a real representation are free from the local gauge anomaly in $8n$
and $4+8n$ dimensions.

Now in $4+8n$~dimensions, this equivalence between real Weyl fermions
and Majorana fermions holds in \emph{euclidean} lattice gauge
theory. Furthermore, these fermions are free from any pathologies even
in a non-perturbative level (this is consistent with the fact that
there is no global gauge anomaly for real representations in
$4+8n$~dimensions~\cite{Okubo:1989vn}). These facts can be shown by
repeating the argument of~ref.~\cite{Suzuki:2000ku}.

How is the situation in $8n$~dimensions? It is natural to expect that
any sensible non-perturbative formulation preserves the above
equivalence between Majorana and real Weyl, because it is a basic
property of the target theory.  Then we find a trouble. In
$8n$~dimensions, it has been known that Weyl fermions in a \emph{real}
representation generally exhibit the global gauge anomaly
(\emph{pseudo-real} representations, on the other hand, are free from
the global gauge anomaly~\cite{Okubo:1989vn}). For example, in
8~dimensions, the Weyl fermion belonging to the {\bf 20'}
representation of the gauge-group~$\SU(4)$ exhibits the global gauge
anomaly ($\pi_8[\SU(4)]=Z_{4!}$). Similarly, in 16~dimensions, the
Weyl fermion in the {\bf 70} representation of $\SU(8)$ is anomalous
($\pi_{16}[\SU(8)]=Z_{8!}$)~\cite{Holman:ef}. So if it was possible to
obtain a simple formulation of the Majorana fermion in $8n$~dimensions
such as eq.~(\ref{twoxten}) (note that this formulation does not refer
to a particular real representation), then either the equivalence with
real Weyl fermions is lost or the global gauge anomaly is not
reproduced. From this argument, we regard a difficulty of the Majorana
decomposition in $8n$~dimensions is a reflection of the global gauge
anomaly.

Next, we consider $d=1+8n$ cases. We note that the Majorana fermion in
$1+8n$~dimensions, after the dimensional reduction, becomes the
Majorana fermion in $8n$~dimensions. This is expected even on the
lattice; if it is possible to construct a lattice action for the
Majorana fermion in $1+8n$~dimensions, then the dimensional reduction
explained in~section~\ref{section5.2} will provide a lattice action of
the Majorana fermion in $8n$~dimensions. However, we argued above that
such a simple action of the Majorana fermion in $8n$~dimensions is
quite unlikely. So, $d=1+8n$ cases are also expected to be
difficult. Similarly, the dimensional reduction of the Majorana-Weyl
fermion in $2+8n$~dimensions will give rise to the Majorana fermion in
$1+8n$~dimensions. So again we expect a difficulty.\footnote{It is
  possible that Majorana-Weyl fermions in~$2+8n$ dimensions themselves
  suffer from the global gauge anomaly, although we do not know of an
  example. Real Weyl fermions in $2+8n$~dimensions are known to be
  free from the global gauge anomaly, when being free from the local
  gauge anomaly~\cite{Okubo:1987dj}.}

In this way, through a ladder of dimensional reductions, three cases,
the Majorana-Weyl fermion in $2+8n$~dimensions, the Majorana fermion
in $1+8n$~dimensions and the Majorana fermion in $8n$~dimensions, are
related.  Then by assuming an equivalence of the Majorana fermion and
the Weyl fermion in a real representation in $8n$~dimensions,
difficulties we found in this paper are related to the global gauge
anomaly. This argument indicates that a lattice formulation of these
fermions is extremely difficult to find. Since the gaugino field of
the $N=1$ super Yang-Mills theory in these dimensions is precisely
represented by these fermions,\footnote{Here, we define a number of
  supercharges~$N$ by taking the smallest irreducible spinor
  representation as a unit. In this convention, the supercharge of the
  $N=1$ super Yang-Mills theory in $2+8n$~dimensions is a single
  Majorana-Weyl fermion.} our argument suggests also that a lattice
formulation of the $N=1$ super Yang-Mills theory in these dimensions
is extremely difficult to find for this reason (besides other
well-known reasons for supersymmetric theories on the
lattice).\footnote{A possible way out for the $N=1$ super Yang-Mills
  theory in $8n$~dimensions is to formulate the gaugino field as the
  lattice Weyl fermion. For this case, however, the transformation
  parameter of the supersymmetry transformation will be a subject of
  the constraint~(\ref{threexfour}) that depends on the gauge field.}

Finally, here is a list of further study. It will be possible to
generalize the argument of~ref.~\cite{Bar:2000qa} to $8n$~dimensions
and study the global gauge anomaly of real Weyl fermions in these
dimensions. (For \emph{pseudo-real} Weyl fermions, an argument similar
to that of~ref.~\cite{Suzuki:2000ku} will be applied.) It will also be
interesting to investigate a possible non-perturbative pathology of
Majorana fermions in $8n$ and $1+8n$~dimensions by using lattice gauge
theory, although we do not know of any pathologies in the latter odd
dimensions in continuum theory.\footnote{It is interesting to note
  that the Majorana fermion in $1+8n$~dimensions suffers from the
  global \emph{gravitational} anomaly~\cite{Alvarez-Gaume:1983ig}.} An
analysis of Majorana-Weyl fermions in $2+8n$~dimensions will be
difficult because a reality of the gauge-group representation does not
imply an absence of the local gauge anomaly. It might also be possible
to establish a no-go theorem for lattice Majorana and Majorana-Weyl
fermions in these dimensions, that is analogous to the
Nielsen--Ninomiya theorem~\cite{Nielsen:1980rz}.\footnote{In this
  respect, it is instructive to see what happens if one adopts the
  action
\begin{equation}
   S=\int\rmd^dx\,{1\over2}\chi^T(x)B_1D_{\rm W}\chi(x)\,,
\end{equation}
for $d=8n$ or~$d=1+8n$, despite the fact that $(B_1D_{\rm
  W})^T\neq-B_1D_{\rm W}$. In this case, one has
\begin{eqnarray}
   S&=&\int\rmd^dx\,{1\over4}\chi^T(x)[B_1D_{\rm W}-(B_1D_{\rm W})^T]\chi(x)
\nonumber\\
   &=&\int\rmd^dx\,{1\over4}\chi^T(x)B_1
   \gamma_\mu(\nabla_\mu+\nabla_\mu^*)\chi(x)\,,
\end{eqnarray}
and the species doubling occurs. A similar conclusion is obtained for
the overlap-Dirac operator.}  We hope to come back some of these
  problems in the near future.

\acknowledgments

H.S.\ would like to thank Takanori Fujiwara, Yoshio Kikukawa and
Kiyoshi Okuyama for helpful discussions.

\appendix

\section{Representation of the Clifford algebra in $d$~dimensional euclidean space}
\label{section7}

Here we summarize our conventions for gamma matrices. Our gamma
matrices are hermitiean and satisfy the Clifford algebra
\begin{equation}
   \{\gamma_\mu,\gamma_\nu\}=2\delta_{\mu\nu}\,,\qquad\mu,\nu=0,1,\ldots,d-1\,.
\label{axone}
\end{equation}

We start with cases of even ($d=2k$) dimensional space. In even
dimensions, the chiral matrix can be defined by
\begin{equation}
   \gamma=i^{-k}\gamma_0\gamma_1\cdots\gamma_{2k-1}\,,
\label{axtwo}
\end{equation}
such that
\begin{equation}
   \{\gamma,\gamma_\mu\}=0,\qquad\gamma^\dagger=\gamma\,,\qquad\gamma^2=1\,.
\label{axthree}
\end{equation}
Since $\pm\gamma_\mu^*$ obey the Clifford algebra identical to
eq.~(\ref{axone}), $\gamma_\mu$ and $\pm\gamma_\mu^*$ are related by
similarity transformations
\begin{equation}
   B_1\gamma_\mu B_1^{-1}=(-1)^k\gamma_\mu^*=(-1)^k\gamma_\mu^T\,,\qquad
   B_2\gamma_\mu B_2^{-1}=(-1)^{k+1}\gamma_\mu^*=(-1)^{k+1}\gamma_\mu^T\,.
\label{axfour}
\end{equation}
One can choose the phase of these matrices such that
\begin{equation}
   B_1^{-1}=B_1^\dagger\,,\qquad B_2^{-1}=B_2^\dagger\,.
\label{axfive}
\end{equation}
These $B_1$ and~$B_2$ are referred to as charge conjugation matrices
in this paper. Note that
\begin{equation}
   B_1\gamma B_1^{-1}=B_2\gamma B_2^{-1}
   =(-1)^k\gamma^*=(-1)^k\gamma^T\,.
\label{axsix}
\end{equation}
We adopt a representation of gamma matrices such that\footnote{For
  example, we can take
\begin{eqnarray}
   \gamma_0&=&\sigma_1\otimes\sigma_3\otimes\sigma_3\otimes\cdots
   \otimes\sigma_3\otimes\sigma_3\,,\qquad
   \gamma_1=\sigma_2\otimes\sigma_3\otimes\sigma_3\otimes\cdots
   \otimes\sigma_3\otimes\sigma_3\,,
\nonumber\\
   \gamma_2&=&1\otimes\sigma_1\otimes\sigma_3\otimes\cdots
   \otimes\sigma_3\otimes\sigma_3\,,\qquad
   \gamma_3=1\otimes\sigma_2\otimes\sigma_3\otimes\cdots
   \otimes\sigma_3\otimes\sigma_3\,,
\nonumber\\
   &&\qquad\qquad\qquad\qquad\qquad\qquad\quad\vdots
\nonumber\\
   \gamma_{2k-2}&=&1\otimes1\otimes1\otimes\cdots
   \otimes1\otimes\sigma_1\,,\qquad
   \gamma_{2k-1}=1\otimes1\otimes1\otimes\cdots
   \otimes1\otimes\sigma_2\,,\qquad
\end{eqnarray}
and
\begin{equation}
   \gamma=\sigma_3\otimes\cdots \otimes\sigma_3\,.
\end{equation}}
\begin{eqnarray}
   &&\gamma_0,\gamma_2,\ldots,\gamma_{2k-2}:\hbox{real}\,,
\nonumber\\
   &&\gamma_1,\gamma_3,\ldots,\gamma_{2k-1}:\hbox{purely imaginary}\,.
\label{axnine}
\end{eqnarray}
In this representation, the chiral matrix is real,
$\gamma^*=\gamma^T=\gamma$.  Charge conjugation matrices in our
representation are then given by
\begin{equation}
   B_1=\gamma_1\gamma_3\ldots\gamma_{2k-1}\,,\qquad B_2=\gamma B_1\,.
\label{axten}
\end{equation}
With a help of these explicit forms, we find
\begin{eqnarray}
   B_1^T&=&(-1)^{k(k+1)/2}B_1\,,\qquad B_2^T=(-1)^{k(k-1)/2}B_2\,,
\label{axeleven}\\
   B_1^*B_1&=&(-1)^{k(k+1)/2}\,,\qquad B_2^*B_2=(-1)^{k(k-1)/2}\,,
\label{axtwelve}
\end{eqnarray}
but actually these relations hold irrespective of the
representation.\footnote{Under a change of the representation,
  $\gamma_\mu\to U^{-1}\gamma_\mu U$, where $U$~is a unitary matrix, a
  charge conjugation matrix~$B$ changes as~$B\to U^TBU$.} As relations
particularly hold in our representation, we have
\begin{eqnarray}
   B_1^*&=&(-1)^kB_1\,,\qquad B_2^*=(-1)^kB_2\,,
\label{axthirteen}\\
   B_1^2&=&(-1)^{k(k-1)/2}\,,\qquad B_2^2=(-1)^{k(k+1)/2}\,.
\label{axfourteen}
\end{eqnarray}

For odd ($d=2k+1$) dimensional space, gamma matrices are obtained from
those of $2k$ dimensions by
\begin{equation}
   \gamma_\mu\,,\quad\mu=0,1,\ldots,2k-1\,\qquad\gamma_{2k}=\gamma\,.
\label{axfifteen}
\end{equation}
From eqs.~(\ref{axfour}) and~(\ref{axsix}), one sees that only $B_1$
is the charge conjugation matrix such that
\begin{equation}
   B_1\gamma_\mu B_1^{-1}=(-1)^k\gamma_\mu^*=(-1)^k\gamma_\mu^T\,,\qquad
   \mu=0,1,\ldots,2k\,.
\label{axsixteen}
\end{equation}

\section{Charge conjugation properties of lattice Dirac operators}\label{section8}

The lattice covariant differences are defined by
\begin{eqnarray}
   \nabla_\mu\psi(x)&=&{1\over a}
   \{R[U(x,\mu)]\psi(x+a\hat\mu)-\psi(x)\}\,,
\nonumber\\
   \nabla_\mu^*\psi(x)&=&{1\over a}
   \{\psi(x)-R[U(x-a\hat\mu,\mu)]^{-1}\psi(x-a\hat\mu)\}\,,
\label{bxone}
\end{eqnarray}
where $U(x,\mu)$ denotes a link variable and $R$ is the gauge-group
representation to which the fermion belongs; $\hat\mu$ is the unit
vector in the direction~$\mu$. We assume that $R$ is a real
representation in what follows. The Wilson-Dirac operator is defined
by
\begin{equation}
   D_{\rm W}={1\over2}[\gamma_\mu(\nabla_\mu^*+\nabla_\mu)
     -a\nabla_\mu^*\nabla_\mu]\,.
\label{bxtwo}
\end{equation}
From the relation~$\nabla_\mu^T=-\nabla_\mu^*$ and formulas in
appendix~\ref{section7}, we find
\begin{equation}
   D_{\rm W}^T=\cases{
   B_2D_{\rm W}B_2^{-1}\,,\quad B_2^T=+B_2\,,&for $d=8n$,\cr
   \hbox{no simple relation}&for $d=1+8n$,\cr
   B_1D_{\rm W}B_1^{-1}\,,\quad B_1^T=-B_1\,,&for $d=2+8n$,\cr
   B_1D_{\rm W}B_1^{-1}\,,\quad B_1^T=-B_1\,,&for $d=3+8n$,\cr
   B_2D_{\rm W}B_2^{-1}\,,\quad B_2^T=-B_2\,,&for $d=4+8n$,\cr}
\label{bxthree}
\end{equation}
and
\begin{equation}
   D_{\rm W}^*=\cases{
   B_1D_{\rm W}B_1^{-1}\,,\quad B_1^T=+B_1\,,&for $d=8n$,\cr
   B_1D_{\rm W}B_1^{-1}\,,\quad B_1^T=+B_1\,,&for $d=1+8n$,\cr
   B_2D_{\rm W}B_2^{-1}\,,\quad B_2^T=+B_2\,,&for $d=2+8n$,\cr
   \hbox{no simple relation}\,,&for $d=3+8n$,\cr
   B_1D_{\rm W}B_1^{-1}\,,\quad B_1^T=-B_1,&for $d=4+8n$.\cr}
\label{bxfour}
\end{equation}

The overlap-Dirac operator is defined by
\begin{equation}
   D={1\over a}[1-A(A^\dagger A)^{-1/2}]\,,\qquad A={1\over a}-D_{\rm
     W}\,,
\label{bxfive}
\end{equation}
from the Wilson-Dirac operator~$D_{\rm W}$. For this operator, we have
\begin{equation}
   D^T=\cases{
   B_2DB_2^{-1}\,,\quad B_2^T=+B_2\,,&for $d=8n$,\cr
   \hbox{no simple relation}&for $d=1+8n$,\cr
   B_1DB_1^{-1}\,,\quad B_1^T=-B_1\,,&for $d=2+8n$,\cr
   \hbox{no simple relation}&for $d=3+8n$,\cr
   B_2DB_2^{-1},\quad B_2^T=-B_2,&for $d=4+8n$.\cr}
\label{bxsix}
\end{equation}

\section{Positivity of the Majorana pfaffian in $2+8n$ and~$4+8n$ dimensions}\label{section9}

In this appendix, we demonstrate that the pfaffian~(\ref{twoxten}) for
$d=2+8n$ and for~$d=4+8n$ is a non-negative function of a gauge-field
configuration, up to a proportionality constant. To show this, we use
the eigenfunctions of the hermitean operator~$H$~(\ref{twoxthirteen}):
\begin{equation}
   H\varphi_n(x)=\lambda_n\varphi_n(x)\,,\qquad
   (\varphi_m,\varphi_n)=\delta_{m,n}\,.
\label{cxone}
\end{equation}
Also we recall the lattice index theorem~\cite{Hasenfratz:1998ft}
\begin{equation}
   \Tr\Gammait=n_+-n_-\,,
\label{cxtwo}
\end{equation}
where the operator~$\Gammait$ is defined by eq.~(\ref{twoxfourteen});
$n_+$ and~$n_-$ denote a number of zero modes of positive and negative
chiralities, respectively. This relation can be shown by noting that
the mode~$\Gammait\varphi_n$ is orthogonal to~$\varphi_n$
for~$\lambda_n\neq0$, because it \pagebreak[3] has an opposite-sign
eigenvalue~$-\lambda_n$ due to eq.~(\ref{twoxfifteen}). One can also
show the chirality sum-rule~\cite{Chiu:1998bh}
\begin{equation}
   \Tr\gamma=n_+-n_-+N_+-N_-=0\,,
\label{cxthree}
\end{equation}
where $N_+$ and~$N_-$ denote a number of modes such that
$H\varphi_n=+(2/a)\varphi_n$ and~$H\varphi_n=-(2/a)\varphi_n$,
respectively.

We start with the case of~$d=2+8n$. First we note that for~$d=2+8n$,
the transpose of~$\Gammait$ is given by $\Gammait^T=-B_2\Gammait
B_2^{-1}$ and the index vanishes
\begin{equation}
   n_+-n_-=\Tr\Gammait=-\Tr\Gammait=0\,.
\label{cxfour}
\end{equation}
The Dirac fermion in a real representation is thus insensitive to a
difference of topological sectors; all possible zero modes are not
topological but accidental. Having this fact in mind, we evaluate the
functional integral
\begin{equation}
   \langle1\rangle_{\rm M}=\int\prod_x\rmd\chi(x)\,
   \exp\left[-a^d\sum_x\,{1\over2}\chi^T(x)B_1D\chi(x)\right],
\label{cxfive}
\end{equation}
by expanding the field~$\chi(x)$ by a complete system of functions. We
first take eigenfunctions~$\varphi_n$ in eq.~(\ref{cxone}) with
\emph{positive} eigenvalues~$\lambda_n>0$ only. From these functions,
we can construct eigenfunctions with \emph{negative} eigenvalues
by~$B_2^{-1}\varphi_n^*$,
\begin{equation}
   HB_2^{-1}\varphi_n^*(x)=-\lambda_nB_2^{-1}\varphi_n^*(x)\,,\qquad
   \lambda_n>0\,.
\label{cxsix}
\end{equation}
Since $B_2^*B_2=1$, $\{\varphi_n,B_2^{-1}\varphi_n^*\}$ form a
complete system.  We thus expand the field in terms of these
eigenfunctions
\begin{equation}
   \chi(x)=\sum_{\lambda_n>0}[\varphi_n(x)c_n+B_2^{-1}\varphi_n^*(x)b_n]\,.
\label{cxseven}
\end{equation}
Then by noting~$B_1D=-B_2H$ and $(\varphi_m,B_2^{-1}\varphi_n^*)=0$,
the action becomes
\begin{equation}
   -a^d\sum_x\,{1\over2}\chi^T(x)B_1D\chi(x)
   =\sum_{\lambda_n>0}\lambda_nb_nc_n\,.
\label{cxeight}
\end{equation}
For the integration measure, we have $\prod_x\rmd\chi(x)=
\det^{-1}U\prod_{n=1}^{(1/2)\Tr1}\rmd c_n\prod_{n=1}^{(1/2)\Tr1}\rmd
b_n$ under the change of variables, where the matrix~$U$ is defined by
\begin{equation}
   U_{x,n}=[\varphi_n(x),B_2^{-1}\varphi_n^*(x)]\,.
\label{cxnine}
\end{equation}
For this matrix, one has the relation
\begin{equation}
   \sum_{x,l}U_{m,x}^TB_2U_{x,l}
   \left[\matrix{0&\delta_{l,n}\cr\delta_{l,n}&0\cr}\right]
   =\left[\matrix{\delta_{m,n}&0\cr0&\delta_{m,n}\cr}\right],
\label{cxten}
\end{equation}
from the orthonormality of~$\varphi_n$ and thus
\begin{equation}
   \det\nolimits^{-2}U=(-1)^{[(1/2)\Tr1]^2}\det B_2\,.
\label{cxeleven}
\end{equation}
In this way, we have
\begin{equation}
   \langle1\rangle_{\rm M}=\pm
   e^{i\alpha/2}\prod_{\lambda_n>0}\lambda_n\,, \qquad
   e^{i\alpha}=(-1)^{[(1/2)\Tr1]^2}\det B_2\,.
\label{cxtwelve}
\end{equation}
The over-all $\pm$~sign in~eq.~(\ref{cxtwelve}) has to be chosen such
that $\langle1\rangle_{\rm M}$ changes smoothly when a certain
eigenvalue crosses zero (otherwise there may be a cusp; recall that
eigenvalues come in pair as $\lambda_n$ and~$-\lambda_n$), because we
\emph{know} that $\langle1\rangle_{\rm M}=\Pf(B_1D)$ is definitely a
smooth function of a gauge-field configuration.

Fortunately, it turns out that this artificial sign flip is not
necessary.\footnote{This is not the case for the Wilson-Dirac
  operator. One has to flip the $\pm$~sign in~eq.~(\ref{cxtwelve}) at
  each time when an eigenvalue crosses zero. As the result, the
  pfaffian is not necessarily non-negative.} To see this, we note the
fact that eigenvalues~$\lambda_n\neq2/a$ have a double degeneracy
(this follows from a good chiral property of the overlap-Dirac
operator). Namely, the mode
\begin{equation}
   \phi_n(x)={1\over\sqrt{1-a^2\lambda_n^2/4}}\Gammait
   B_2^{-1}\varphi_n^*(x)\,,
\label{cxthirteen}
\end{equation}
has an identical eigenvalue~$\lambda_n$ to~$\varphi_n$. This
mode~$\phi_n$ is linearly independent from~$\varphi_n$ because
$(\varphi_n,\phi_n)=0$ holds due to~$(\Gammait B_2^{-1})^T=-\Gammait
B_2^{-1}$. From this fact, we have
\begin{equation}
   \langle1\rangle_{\rm M}
   =e^{i\alpha/2}(2/a)^{N_+}\prod_{0<\lambda_n\neq2/a}\lambda_n^2\,,
\label{cxeleven1}
\end{equation}
where we have taken the $+$~sign in~eq.~(\ref{cxtwelve}) for all
gauge-field configurations, because then the expression becomes a
smooth function. The product~$\prod_{\lambda_n}$ in this expression is
understood to be omitting the above double degeneracy, i.e., one
factor~$\lambda_n^2$ for each pair of~$\varphi_n$ and~$\phi_n$. This
expression proves that~$\langle1\rangle_{\rm M}$ is, up to a
proportionality constant, a non-negative function of a gauge-field
configuration.

Next, we consider the case of~$d=4+8n$. The following analysis is
basically identical to that of ref.~\cite{Suzuki:2000ku}. For this
case, the index~$n_+-n_-$ does not vanish in general and there may
exist topological zero modes. To avoid a complication associated to
these zero modes, it is convenient to introduce a mass
term~\cite{Suzuki:2000ku}. Thus, we consider
\begin{equation}
   \langle1\rangle_{\rm M}=\int\prod_x\rmd\chi(x)\,
   \exp\left\{-a^d\sum_x\,\left[{1\over2}\chi^T(x)B_2D\chi(x)
   +{1\over2}im\chi^T(x)B_1\chi(x)
   \right]\right\}.
\end{equation}
For the present case, we use all the eigenfunctions~$\varphi_n$
in~eq.~(\ref{cxone}) to expand the field. However, there is a double
degeneracy, namely, the mode
\begin{equation}
   \phi_n(x)=B_1^{-1}\varphi_n^*(x)\,,
\end{equation}
has the eigenvalue \emph{identical} to that of the
mode~$\varphi_n$. This is linearly independent from~$\varphi_n$,
because $(\varphi_n,\phi_n)=0$ holds (note~$B_1^T=-B_1$). In
particular, $n_\pm$ and~$N_\pm$ are even numbers.  Noting these facts,
we expand the field as
\begin{equation}
   \chi(x)=\sum_{\lambda_n}[\varphi_n(x)c_n+B_1^{-1}\varphi_n^*(x)b_n]\,.
\end{equation}
Then by using $B_2D=B_1H$, the action becomes
\begin{equation}
   -a^d\sum_x\,\left[{1\over2}\chi^T(x)B_2D\chi(x)
   +{1\over2}im\chi^T(x)B_1\chi(x)\right]
   =\sum_{\lambda_n}(\lambda_n+im)b_nc_n\,.
\end{equation}
For the integration measure, we have $\prod_x\rmd\chi(x)=
\det^{-1}V\prod_{n=1}^{(1/2)\Tr1}\rmd c_n\prod_{n=1}^{(1/2)\Tr1}\rmd
b_n$, where the matrix~$V$ is defined by
\begin{equation}
   V_{x,n}=[\varphi_n(x),B_1^{-1}\varphi_n^*(x)]\,,
\end{equation}
and, corresponding to eq.~(\ref{cxten}),
\begin{equation}
   \sum_{x,l}V_{m,x}^TB_1V_{x,l}
   \left[\matrix{0&-\delta_{l,n}\cr\delta_{l,n}&0\cr}\right]
   =\left[\matrix{\delta_{m,n}&0\cr0&\delta_{m,n}\cr}\right],
\end{equation}
and
\begin{equation}
   \det\nolimits^{-2}V=(-1)^{[(1/2)\Tr1]^2}(-1)^{(1/2)\Tr1}\det B_2\,.
\end{equation}
After the integration over $c_n$ and~$b_n$, we thus have
\begin{equation}
   \langle1\rangle_{\rm M}=\pm e^{i\beta/2}(-1)^{(1/4)\Tr1}
   \prod_{\lambda_n}(\lambda_n+im)\,,\qquad
   e^{i\beta}=(-1)^{[(1/2)\Tr1]^2}\det B_1\,,
\label{cxtwentytwo}
\end{equation}
where the product~$\prod_{\lambda_n}$ is understood to be omitting the
double degeneracy, i.e., one factor~$\lambda_n+im$ for each pair
of~$\varphi_n$ and~$\phi_n$. At this point, we further note that
eigenvalues $\lambda_n\neq0$ and~$\lambda_n\neq\pm2/a$ appear in pair
as~$\lambda_n$ and~$-\lambda_n$, because
\begin{equation}
   {1\over\sqrt{1-a^2\lambda_n^2/4}}\Gammait\varphi_n(x)\,,
\end{equation}
has the eigenvalue~$-\lambda_n$. From these facts, we have
\begin{eqnarray}
   \langle1\rangle_{\rm M}
   &=&e^{i\beta/2}(-1)^{(1/4)\Tr1}(im)^{(n_++n_-)/2}(2/a+im)^{N_+/2}
   (-2/a+im)^{N_-/2}\times
\nonumber\\&&
   \times\,\prod_{0<\lambda_n\neq2/a}(-\lambda_n+im)(\lambda_n+im)\,,
\end{eqnarray}
where we have taken the $+$~sign in~eq.~(\ref{cxtwentytwo}) for all
configurations, because then the expression becomes a smooth
function. Then from a counting of a number of eigenfunctions, one
finds that a number of terms in the last product
is~$(\Tr1-N_+-N_--n_+-n_-)/4$. Therefore, we finally have
\begin{eqnarray}
   \langle1\rangle_{\rm M}&=&e^{i\beta/2}(-1)^{(n_+-n_-)/4}
   (2/a-im)^{(n_+-n_-)/2}\times
\nonumber\\&&
   \times\, m^{(n_++n_-)/2}(4/a^2+m^2)^{N_+/2}
   \prod_{0<\lambda_n\neq2/a}(\lambda_n^2+m^2)\,,
\end{eqnarray}
where a use of the chirality sum-rule~(\ref{cxthree}) has been
made. It is interesting to note that the
prefactor~$(-1)^{(n_+-n_-)/4}(2/a-im)^{(n_+-n_-)/2}$ in this
expression depends only on a topological sector concerned through the
index~$n_+-n_-$. Assuming that there is no zero modes, this expression
reduces to
\begin{equation}
   \langle1\rangle_{\rm M}=e^{i\beta/2}
   (2/a)^{N_+}\prod_{0<\lambda_n\neq2/a}\lambda_n^2\,,
\end{equation}
in the massless limit. This proves that the pfaffian is, up to a
proportionality constant, a non-negative function of a gauge-field
configuration.


\begin{thebibliography}{99}

\bibitem{Wilson:1975id}
K.G.~Wilson, \emph{Quarks and strings on a lattice}, in \emph{New
  phenomena in subnuclear physics. Part A}, A.~Zichichi ed., Plenum
Press, New York, 1977, p.~69, proceedings of the first half of the
1975 International School of Subnuclear Physics, Erice, Sicily, July
11-August 1, 1975.

\bibitem{Neuberger:1998fp}
H.~Neuberger, \emph{Exactly massless quarks on the lattice},
\plb{417}{1998}{141} [\heplat{9707022}];
\emph{More about exactly massless quarks on the lattice},
\plb{427}{1998}{353} [\heplat{9801031}].

\bibitem{Hasenfratz:1998ft}
P.~Hasenfratz, \emph{Prospects for perfect actions},
\npps{63}{1998}{53} [\heplat{9709110}];
\emph{Lattice QCD without tuning, mixing and current renormalization},
\npb{525}{1998}{401} [\heplat{9802007}];\\
P.~Hasenfratz, V.~Laliena and F.~Niedermayer,
\emph{The index theorem in QCD with a finite cut-off},
\plb{427}{1998}{125} [\heplat{9801021}].

\bibitem{Nicolai:1978vc}
H.~Nicolai, \emph{A possible constructive approach to (super-$\phi^3$)
  in four-dimensions. (I). Euclidean formulation of the model},
\npb{140}{1978}{294};
\emph{A possible constructive approach to super-$\phi^3$ in
  four-dimensions. (II). Regularization of the model},
  \npb{156}{1979}{157};
\emph{A possible constructive approach to (super-$\phi^3$) in
four-dimensions. (III). On the normalization of schwinger functions},
\npb{156}{1979}{177}.

\bibitem{vanNieuwenhuizen:1996tv}
P.~van Nieuwenhuizen and A.~Waldron, \emph{On euclidean spinors and
  Wick rotations}, \plb{389}{1996}{29} [\hepth{9608174}], and
  references therein.

\bibitem{Fujikawa:2001ka}
K.~Fujikawa and M.~Ishibashi, \emph{Lattice chiral symmetry and the
  Wess-Zumino model}, \npb{622}{2002}{115} [\hepth{0109156}];
\emph{Lattice chiral symmetry, Yukawa couplings and the Majorana
  condition}, \plb{528}{2002}{295} [\heplat{0112050}].

\bibitem{Fujikawa:2002is}
K.~Fujikawa, M.~Ishibashi and H.~Suzuki, \emph{Ginsparg-Wilson
  operators and a no-go theorem}, \plb{538}{2002}{197}
[\heplat{0202017}].

\bibitem{Hasenfratz:2001bz}
P.~Hasenfratz, \emph{Lattice 2001: reflections}, \npps{106}{2002}{159}
[\heplat{0111023}].

\bibitem{Fujikawa:2002vj}
K.~Fujikawa, M.~Ishibashi and H.~Suzuki, \emph{CP breaking in lattice
  chiral gauge theories}, \jhep{04}{2002}{046} [\heplat{0203016}];\\
K.~Fujikawa and H.~Suzuki, \emph{Domain wall fermion and CP symmetry
  breaking}, \prd{67}{2003}{034506} [\heplat{0210013}].

\bibitem{Kikukawa:1997qh}
Y.~Kikukawa and H.~Neuberger, \emph{Overlap in odd dimensions},
\npb{513}{1998}{735} [\heplat{9707016}].

\bibitem{Narayanan:1993wx}
R.~Narayanan and H.~Neuberger, \emph{Infinitely many regulator fields
  for chiral fermions}, \plb{302}{1993}{62} [\heplat{9212019}];
\emph{Chiral determinant as an overlap of two vacua},
\npb{412}{1994}{574} [\heplat{9307006}];
\emph{Chiral fermions on the lattice}, \prl{71}{1993}{3251}
     [\heplat{9308011}];
\emph{A construction of lattice chiral gauge theories},
\npb{443}{1995}{305} [\hepth{9411108}];\\
S.~Randjbar-Daemi and J.~Strathdee, \emph{On the overlap formulation
  of chiral gauge theory}, \plb{348}{1995}{543} [\hepth{9412165}];
\emph{Chiral fermions on the lattice}, \npb{443}{1995}{386}
     [\heplat{9501027}];
\emph{On the overlap prescription for lattice regularization of chiral
  fermions}, \npb{466}{1996}{335} [\hepth{9512112}];
\emph{Consistent and covariant anomalies in the overlap formulation of
  chiral gauge theories}, \plb{402}{1997}{134} [\hepth{9703092}].

\bibitem{Huet:1996pw}
P.Y.~Huet, R.~Narayanan and H.~Neuberger, \emph{Overlap formulation of
  Majorana-Weyl fermions}, \plb{380}{1996}{291} [\hepth{9602176}];\\
R.~Narayanan and H.~Neuberger, \emph{Overlap for Majorana-Weyl
  fermions}, \npps{53}{1997}{658} [\heplat{9607080}].

\bibitem{Maru:1997kh}
N.~Maru and J.~Nishimura, \emph{Lattice formulation of supersymmetric
  Yang-Mills theories without fine-tuning}, \ijmpa{13}{1998}{2841}
[\hepth{9705152}].

\bibitem{Kitsunezaki:1997iu}
N.~Kitsunezaki and J.~Nishimura, \emph{Unitary IIB matrix model and
  the dynamical generation of the space time}, \npb{526}{1998}{351}
[\hepth{9707162}].

\bibitem{Witten:fp}
E.~Witten, \emph{An ${\rm SU}(2)$ anomaly}, \plb{117}{1982}{324}.

\bibitem{Elitzur:1984kr}
S.~Elitzur and V.P.~Nair, \emph{Nonperturbative anomalies in higher
  dimensions}, \npb{243}{1984}{205}.

\bibitem{Ginsparg:1982bj}
P.H.~Ginsparg and K.G.~Wilson, \emph{A remnant of chiral symmetry on
  the lattice}, \prd{25}{1982}{2649}.

\bibitem{Luscher:1998pq}
M.~L\"uscher, \emph{Exact chiral symmetry on the lattice and the
  Ginsparg-Wilson relation}, \plb{428}{1998}{342} [\heplat{9802011}].

\bibitem{Narayanan:1998uu}
R.~Narayanan, \emph{Ginsparg-Wilson relation and the overlap formula},
\prd{58}{1998}{097501} [\heplat{9802018}];\\
F.~Niedermayer, \emph{Exact chiral symmetry, topological charge and
  related topics}, \npps{73}{1999}{105} [\heplat{9810026}].

\bibitem{Hernandez:1999et}
P.~Hern\'andez, K.~Jansen and M.~L\"uscher, \emph{Locality properties
  of Neuberger's lattice Dirac operator}, \npb{552}{1999}{363}
[\heplat{9808010}];\\
H.~Neuberger, \emph{Bounds on the Wilson Dirac operator},
\prd{61}{2000}{085015} [\heplat{9911004}].

\bibitem{Suzuki:2000ku}
H.~Suzuki, \emph{Real representation in chiral gauge theories on the
  lattice}, \jhep{10}{2000}{039} [\heplat{0009036}].

\bibitem{Neuberger:1997bg}
H.~Neuberger, \emph{Vector like gauge theories with almost massless
  fermions on the lattice}, \prd{57}{1998}{5417} [\heplat{9710089}].

\bibitem{Kaplan:1999jn}
D.B.~Kaplan and M.~Schmaltz, \emph{Supersymmetric Yang-Mills theories
  from domain wall fermions}, \newjournal{Chin.\ J.\
  Phys.}{CJOPA}{38}{2000}{543} [\heplat{0002030}].

\bibitem{Feo:2002yi}
For a review see A.~Feo, \emph{Supersymmetry on the lattice},
\npps{119}{2003}{198} [\heplat{0210015}], and references therein.

\bibitem{Nielsen:1980rz}
H.B. Nielsen and M.~Ninomiya, \emph{Absence of neutrinos on a lattice,
  1. Proof by homotopy theory}, \npb{185}{1981}{20}, erratum \ibid{B
  195}{1982}{541};
\emph{No go theorem for regularizing chiral fermions},
\plb{105}{1981}{219};\\
D.~Friedan, \emph{A proof of the Nielsen-Ninomiya theorem},
\cmp{85}{1982}{481}.

\bibitem{Bunk:2004br}
B.~Bunk, M.~Della~Morte, K.~Jansen and F.~Knechtli, \emph{Locality
  with staggered fermions}, \heplat{0403022}.

\bibitem{Kaplan:1992bt}
D.B. Kaplan, \emph{A method for simulating chiral fermions on the
  lattice}, \plb{288}{1992}{342} [\heplat{9206013}].

\bibitem{Luscher:1999du}
M.~L\"uscher, \emph{Abelian chiral gauge theories on the lattice with
  exact gauge invariance}, \npb{549}{1999}{295} [\heplat{9811032}].

\bibitem{Luscher:1999un}
M.~L\"uscher, \emph{Weyl fermions on the lattice and the non-abelian
  gauge anomaly}, \npb{568}{2000}{162} [\heplat{9904009}].

\bibitem{Suzuki:2000ii}
H.~Suzuki, \emph{Anomaly cancellation condition in lattice gauge
  theory}, \npb{585}{2000}{471} [\heplat{0002009}];\\
H.~Igarashi, K.~Okuyama and H.~Suzuki, \emph{Errata and addenda to
  ``Anomaly cancellation condition in lattice gauge theory''},
\heplat{0012018}.

\bibitem{Kikukawa:2000kd}
Y.~Kikukawa and Y.~Nakayama, \emph{Gauge anomaly cancellations in
  ${\rm SU}(2)_L \times {\rm U}(1)_Y$ electroweak theory on the
  lattice}, \npb{597}{2001}{519} [\heplat{0005015}].

\bibitem{Luscher:2000zd}
M.~L\"uscher, \emph{Lattice regularization of chiral gauge theories to
  all orders of perturbation theory}, \jhep{06}{2000}{028}
[\heplat{0006014}].

\bibitem{Kikukawa:2001mw}
Y.~Kikukawa, \emph{Domain wall fermion and chiral gauge theories on
  the lattice with exact gauge invariance}, \prd{65}{2002}{074504}
[\heplat{0105032}].

\bibitem{Kadoh:2003ii}
D.~Kadoh, Y.~Kikukawa and Y.~Nakayama, \emph{Solving the local
  cohomology problem in ${\rm U}(1)$ chiral gauge theories within a finite
  lattice}, \heplat{0309022};\\
D.~Kadoh and Y.~Kikukawa, \emph{A numerical solution to the local
  cohomology problem in ${\rm U}(1)$ chiral gauge theories}, \heplat{0401025}.

\bibitem{Golterman:2004qv}
M.~Golterman and Y.~Shamir, \emph{${\rm SU}(N)$ chiral gauge theories
  on the lattice}, \heplat{0404011}, and references therein.

\bibitem{Redlich:1983kn}
A.N.~Redlich, \emph{Gauge noninvariance and parity nonconservation of
  three- dimensional fermions}, \prl{52}{1984}{18};
\emph{Parity violation and gauge noninvariance of the effective gauge
  field action in three-dimensions}, \prd{29}{1984}{2366};\\
A.J.~Niemi and G.W.~Semenoff, \emph{Axial anomaly induced fermion
  fractionization and effective gauge theory actions in odd
  dimensional space-times}, \prl{51}{1983}{2077}.

\bibitem{So:1984nf}
H.~So, \emph{Induced Chern-Simons class with lattice fermions},
\ptp{73}{1985}{528};
\emph{Induced topological invariants by lattice fermions in odd
  dimensions}, \ptp{74}{1985}{585};\\
A.~Coste and M.~L\"uscher, \emph{Parity anomaly and fermion boson
  transmutation in three-dimensional lattice QED},
\npb{323}{1989}{631}.

\bibitem{Narayanan:1997by}
R.~Narayanan and J.~Nishimura, \emph{Parity-invariant lattice
  regularization of a three- dimensional gauge-fermion system},
\npb{508}{1997}{371} [\hepth{9703109}].

\bibitem{Okubo:1989vn}
S.~Okubo and Y.~Tosa, \emph{Further study of global gauge anomalies of
  simple groups}, \prd{40}{1989}{1925}.

\bibitem{Holman:ef}
R.~Holman and T.W. Kephart, \emph{Global anomalies in Yang-Mills
  theories in higher dimensions}, \plb{167}{1986}{417}.

\bibitem{Okubo:1987dj}
S.~Okubo, H.~Zhang, Y.~Tosa and R.E. Marshak, \emph{${\rm SU}(N)$
  global gauge anomalies in even dimensions}, \prd{37}{1988}{1655}.

\bibitem{Bar:2000qa}
O.~B\"ar and I.~Campos, \emph{Global anomalies in chiral gauge
  theories on the lattice}, \npb{581}{2000}{499} [\heplat{0001025}].

\bibitem{Alvarez-Gaume:1983ig}
L.~Alvarez-Gaum\'e and E.~Witten, \emph{Gravitational anomalies},
\npb{234}{1984}{269}.

\bibitem{Chiu:1998bh}
T.W.~Chiu, \emph{The spectrum and topological charge of exactly
  massless fermions on the lattice}, \prd{58}{1998}{074511}
[\heplat{9804016}];\\
K.~Fujikawa, \emph{Relation ${\rm Tr}\gamma_5 = 0$ and the index
  theorem in lattice gauge theory}, \prd{60}{1999}{074505}
[\heplat{9904007}].

\end{thebibliography}
\end{document}